\documentclass[sort&compress, 3p]{elsarticle}
\usepackage[utf8]{inputenc}
\usepackage{amsmath}
\usepackage{amssymb}
\usepackage{bm}
\usepackage{mathtools}
\usepackage{caption}
\usepackage{subcaption}
\usepackage{pdfpages}
\newcommand{\deriv}[3][]{\ensuremath{\dfrac{\partial^{#1} {#2}}{\partial {#3}^{#1}}}}

\begin{document}
\begin{frontmatter}
\title{Learning constitutive models from microstructural simulations via a non-intrusive reduced basis method}
\author[1]{Theron Guo\corref{cor1}}\ead{t.guo@tue.nl}
\author[2]{Ond\v{r}ej Roko\v{s}}\ead{o.rokos@tue.nl}
\author[1]{Karen Veroy}\ead{k.p.veroy@tue.nl}
\cortext[cor1]{Corresponding author}
\address[1]{Centre for Analysis, Scientific Computing and Applications, Department of Mathematics and Computer Science}
\address[2]{Mechanics of Materials, Department of Mechanical Engineering\\ Eindhoven University of Technology, 5612 AZ Eindhoven, The Netherlands}
\begin{abstract}
In order to optimally design materials, it is crucial to understand the structure-property relations in the material by analyzing the effect of microstructure parameters on the macroscopic properties. In computational homogenization, the microstructure is thus explicitly modeled inside the macrostructure, leading to a coupled two-scale formulation. Unfortunately, the high computational costs of such multiscale simulations often render the solution of design, optimization, or inverse problems infeasible. To address this issue, we propose in this work a non-intrusive reduced basis method to construct inexpensive surrogates for parametrized microscale problems; the method is specifically well-suited for multiscale simulations since the coupled simulation is decoupled into two independent problems:~(1)~solving the microscopic problem for different (loading or material) parameters and learning a surrogate model from the data; and~(2)~solving the macroscopic problem with the learned material model. The proposed method has three key features. First, the microscopic stress field can be fully recovered, which is useful for instance for revealing local stress concentrations inside the microstructure. Second, the method is able to accurately predict the stress field for a wide range of material parameters; furthermore, the derivatives of the effective stress with respect to the material parameters are available and can be readily utilized in solving optimization problems. Finally, it is more data efficient, i.e. requiring less training data, as compared to directly performing a regression on the effective stress. To construct the surrogate model, first, a proper orthogonal decomposition is performed on precomputed microscopic stress field snapshots to find a reduced basis for the stress. Second, a regression is employed to infer the coefficients of the reduced basis approximation for any arbitrary parameter value, thus enabling a rapid online evaluation of the microscopic stress. Equipped with the stress field, the effective stress and its partial derivatives can then be derived analytically. For the microstructures in the two test problems considered, the mean approximation error of the effective stress is as low as 0.1\% despite using a relatively small training dataset. Embedded into the macroscopic problem, the reduced order model leads to an online computational speed up of approximately three orders of magnitude while maintaining a high accuracy as compared to the FE$^2$ solver.

\end{abstract}
\begin{keyword}
Non-intrusive reduced basis method \sep proper orthogonal decomposition \sep computational homogenization \sep Gaussian process regression \sep multiscale analysis
\end{keyword}
\end{frontmatter}
\section{Introduction}
\label{sec:introduction}

Simulation methods are widely utilized to bridge the understanding between macroscopic behavior and the microstructure of the material. This can be achieved by using multiscale methods such as computational homogenization (CH). In CH, the microstructure and macrostructure are both separately modeled, with the microstructure essentially replacing the constitutive model of the macroscopic problem. The microstructure is fully defined on a representative volume element (RVE), which can be highly heterogeneous and exert highly nonlinear behavior, while the macrostructure is then assumed to be homogeneous.

A comprehensive overview on CH is given in Geers et al.~\cite{Geers2010,Geers2016} and Matou\v{s} et al.~\cite{Matous2017}. The two-scale problem can be solved with either a nested finite element (FE) scheme, first proposed in Feyel~\cite{Feyel1999MultiscaleStructures}, or by a combination of the FE method and fast Fourier transform (FFT)~\cite{Moulinec1998AMicrostructure}. More recently, other mesh-free methods based on deep neural networks such as physics-informed neural networks~\cite{Raissi2019, Haghighat2021AMechanics} and the deep collocation method \cite{Weinan2018, Samaniego2020} have also been used for the solution of partial differential equations (PDEs).

Unfortunately, solving the two-scale system in CH is computationally expensive and can quickly exceed available resources, limiting the usage to mostly two-dimensional applications~\cite{Kouznetsova2001, Rokos2019a, Nguyen2012, Xia2017}. Moreover, the difficulty in obtaining the consistent effective stiffness from the microscale simulation presents an additional hurdle, as it is typically required for the macroscopic Newton solver~\cite{Saeb2016}. Popular approaches to find the appropriate stiffness include the Lagrange multiplier method~\cite{Miehe2002, Miehe2003ComputationalEnergy}, condensation method~\cite{Kouznetsova2001}, or the perturbation method~\cite{Temizer2008, Okada2010StudyCost}. However, all these methods introduce additional problems such as increased storage requirements or higher computational costs.

To overcome this computational burden, several different reduced order models have been proposed to speed up or to replace the microscopic simulation. These can be largely classified according to how they treat parameters, and according to whether or not they are intrusive. The ability to handle parameterized problems makes a method suitable for inverse tasks such as parameter identification, material design or optimization. Intrusive methods require the modification of the underlying PDE solver, often making them impractical to use. In contrast, non-intrusive (data-driven) methods can be wrapped around an already existent PDE solver, allowing for a simpler adoption. In the context of CH, data-driven methods typically generate datasets from microscopic simulations and learn an effective constitutive model from the data, essentially decoupling the two-scale problem.

One popular data-driven framework was introduced in Kirchdoerfer et al.~\cite{Kirchdoerfer2016} and extended in \cite{Kirchdoerfer2017DataSets, Eggersmann2019, Karapiperis2021Data-DrivenMechanics}. These works propose a distance minimizing scheme, with which pairs of strain and stress data can be directly utilized inside a simulation without the need of deriving an empirical constitutive model. This dataset can come either from experiments or (microscopic) simulations. However, due to the nature of this method, it is only able to describe and reconstruct a given material dataset and cannot predict new materials. 

Recognizing that stress-strain data is generally difficult to obtain from experiments, a slightly different approach was introduced in Huang et al.~\cite{Huang2020a, Huang2021}, where the constitutive model is approximated by a neural network, which is then inversely trained from load-displacement data. Unfortunately, the trained parameters inside the neural network (weights and biases) do not represent actual physical parameters, and therefore the method can only describe the given dataset.

Several other works attempted to learn a constitutive model directly from pairs of stress and strain data with different neural network architectures~\cite{Ghaboussi1991Knowledge-basedNetworks, Ghaboussi1998, Ghavamian2019, Wu2020a,Logarzo2021, Mozaffar2019, Linka2021ConstitutiveLearning, Masi2021Thermodynamics-basedModeling}. After the pioneering works in Ghaboussi et al.~\cite{Ghaboussi1991Knowledge-basedNetworks}, several works have applied deep neural networks (DNN) to different areas of constitutive modeling. In~\cite{Ghavamian2019, Wu2020a,Logarzo2021} different recurrent neural network architectures were proposed to learn an inelastic material model from stress and strain loading data that were obtained from RVE simulations. However, the models were trained only with deformation data and did not consider parameters, inhibiting their use for finding new materials and designs. Recent works have also attempted to overcome this challenge. In Le et al.~\cite{Le2015}, it was proposed to use neural networks to learn the effective potential of a hyperelastic material from which the effective stress and stiffness can be derived, while also including multiple microstructural parameters. In Mozaffar et al.~\cite{Mozaffar2019} a recurrent neural network that can treat both temporal (strain path) and non-temporal (volume fraction) parameters simultaneously was proposed, and the stress prediction for a class of composite materials was illustrated. More recently, works \cite{Linka2021ConstitutiveLearning,Masi2021Thermodynamics-basedModeling,Huang2021} have also attempted to construct a neural network that learns constitutive models that fulfill the thermodynamical laws as defined in \cite{coleman1963}.

Although these neural network-based methods are capable of finding accurate surrogate models, they typically require huge amounts of data, which are often generated from simulations. For complex microstructures, running these simulations repeatedly might be computational too expensive. Furthermore, these surrogate models generally only consider the effective quantities, which are obtained by averaging the microscopic quantities, and neglect all the field information of the microscopic simulation. This means that the connection between macro- and microstructure is completely lost and local microscopic quantities can no longer be recovered.

Intrusive methods, on the other hand, still consider the microscopic PDE and therefore often require far less data due to the known physics. Instead of replacing the microscopic problem, the solution is accelerated. One notable method is the Transformation Field Analysis (TFA)~\cite{dvorak1992}, which is specifically designed for simulations involving inelastic materials. By considering the field of internal variables to be piecewise uniform, the computational cost of evaluating the nonlinear terms is greatly decreased. Moreover, the number of internal variables that needs to be tracked is also reduced. Later, the method was generalized to non-uniform fields and referred to as Nonuniform Transformation Field Analysis (NTFA)~\cite{Michel2003, Fritzen2013ReducedFormulation}. Even though successful in accelerating two-scale simulations, the reduced model is designed for a specific microscopic material model and therefore cannot handle material parameters~\cite{Fritzen2015}.

Another approach called the Self-consistent Clustering Analysis (SCA) was proposed in Liu et al.~\cite{Liu2016} and extended in~\cite{Yu2019}. The idea of this method is to find material points within a microstructure that exhibit similar deformation behavior and to group them into clusters. It is then assumed that each cluster exhibits the same deformation, therefore reducing the number of independent points to the number of clusters. The compressed problem is then solved with a FFT approach. A similar approach was proposed in~\cite{Wulfinghoff2018}, however, in that work the problem is solved with a three-field Hashin–Shtrikman type variational formulation. Similarly to (N)TFA, this method does not offer a direct way of treating material parameters.

In the Proper Generalized Decomposition (PGD)~\cite{Chinesta2014, Ladeveze2010}, the solution to a PDE is approximated by a linear combination of terms that each consist of a separated functional representation for each independent variable (i.e. coordinates, time, and parameters). The PGD approach has been applied to nonlinear solid mechanics problems, see, e.g.,~\cite{Niroomandi2013}. It has also been combined with the LATIN multiscale method for non-linear problems, see, e.g.,~\cite{Ladeveze2010}. However, the method becomes increasingly more difficult for severe nonlinearities, especially in combination with varying parameters~\cite{Chinesta2014}.

A related approach is the reduced basis (RB) method~\cite{Prudhomme2002ReliableMethods,Quarteroni2015,Hesthaven2016CertifiedEquations} which attempts to reduce the parametric problem by searching for the PDE solution on a reduced space spanned by global parameter-independent basis functions. The basis functions can be constructed by a proper orthogonal decomposition (POD) of a collection of high-fidelity simulations, which are often referred to as snapshots. The solution can then be found by either solving the reduced problem with a (Petrov-)Galerkin projection~\cite{Carlberg2011EfficientApproximations, Radermacher2013, Soldner2017, Ghavamian2017, Ryckelynck2009, Hernandez2014} or as proposed in more recent works by using a regression-based approach to find the basis coefficients~\cite{Guo2018, Hesthaven2018, Kast2020, Swischuk2019, Guenot2013, Yang2020}. Material parameters can naturally be dealt with in this framework.

The projection-based approach works especially well when the problem has an affine dependence on the parameters, allowing the decoupling into an offline (construction of the RB) and online stage (solving the reduced set of equations). However, this is not the case for the general solution of a nonlinear problem since the nonlinear terms will still depend on the full scale problem. To work around this issue, methods such as the empirical interpolation method (EIM)~\cite{Barrault2004} and its discrete variant~\cite{Chaturantabut2010NonlinearInterpolation}, or hyperreduction~\cite{Ryckelynck2009} can be used to approximate the nonlinear term and recover the affine dependence. In the context of CH, POD was first applied on a hyperelastic material in Yvonnet et al.~\cite{Yvonnet2007}. However, in this work the Galerkin approach was directly applied without any special treatment of the nonlinear term, thus resulting in a rather modest speedup. In Hern\'{a}ndez et al.~\cite{Hernandez2014}, the computation of the nonlinear term was accelerated by a hyperreduction scheme and an algorithm to find appropriate integration points was proposed. In Soldner et al.~\cite{Soldner2017}, POD was combined with three different hyperreduction schemes and it was found that some approximations led to convergence problems in some cases. Another problem is that there does not exist a practical way of calculating the consistent effective stiffness~\cite{Yvonnet2007}, as this would require the full assembly of the microscopic tangent matrix. If the tangent matrix was approximated with a hyperreduction scheme, its symmetry might not be preserved and therefore an inconsistent effective stiffness obtained with a perturbation method might be preferable~\cite{Soldner2017}.

Instead of solving the reduced system, the regression-based approach infers the coefficients of the corresponding RB functions from a regression. In doing so, the online phase only consists of evaluating the coefficients from which the solution field can be directly obtained. Furthermore, the method becomes non-intrusive, similar to the other data-driven approaches. However, as observed in Hesthaven et al.~\cite{Hesthaven2018}, considerably more snapshots might be needed to construct an accurate regression as compared to an accurate basis. For that reason, active learning algorithms~\cite{Guo2018, Yang2020} or a bifidelity reconstruction~\cite{Kast2020} were recently proposed in order to minimize the number of snapshots needed.

While the projection-based method efficiently approximates only the field variable (here, the displacements), the regression-based approach could, in principle, be applied to any derived variable obtained in the high-fidelity solution (here, e.g., the stress). In~\cite{Swischuk2019}, Swischuk et al. utilized a regression-based POD to directly find the strain field. In the case of CH, the microscopic displacements are not needed, but only the effective stress is required. Hence, in this work, we propose a regression-based RB surrogate model for the microscopic stress field, which is specifically geared towards multiscale simulations, to combine the advantages of intrusive and non-intrusive methods: (i) Due to the non-intrusiveness, this method can be easily implemented and the two-scale problem is decoupled into two independent problems, (ii) the approximation of the stress field allows a direct way of computing the effective stress and its derivative with respect to any (loading or material) parameter and (iii) even though the method requires more data as compared to the projection-based POD, it is more data efficient as compared to the pure data-driven methods. Furthermore, there are two reasons that favor the stress-based regression over the usual displacement-based approach:
\begin{enumerate}
    \item If the displacements were approximated, the stresses would still need to be computed from the displacements. This means that the material model must be implemented again, or the displacement data has to be fed into the microstructure solver to return the stresses, thus negating the advantages of the non-intrusive method. Furthermore this would lead to unnecessary additional computation, leading to decreased efficiency. 
    \item Inelastic material models are defined in stress rate rather than stresses, including internal variables in their formulation that need to be tracked. It would thus not be possible to determine the stresses using only the displacement field since the evolution of the internal variables would not be known. As a result, the procedure would become highly inefficient or even inapplicable to history-dependent material behavior. On the other hand, works in DNN have shown that history-dependent behavior can be learned from stress and strain data alone without the need for internal variables~\cite{Mozaffar2019, Abueidda2021, Wu2020a, Ghaboussi1991Knowledge-basedNetworks, Ghaboussi1998, Ghavamian2019, Logarzo2021}.
\end{enumerate}

The remainder of this paper is organized as follows. In Section \ref{sec:theory}, the two-scale theory of first-order computational homogenization is reviewed. Following that, Section \ref{sec:reducedbasis} presents how the non-intrusive reduced order model is constructed. Moreover, an error estimation and a comparison to neural networks is provided. Two microstructural simulations and a two-scale problem are addressed in Section \ref{sec:results}, testing the capabilities of the proposed framework. A summary of the findings and concluding remarks are presented in Section \ref{sec:conclusion}.

\paragraph{Notation} In this work, italic bold symbols are employed for coordinates and functions, such as the coordinates $\bm{X}$, the displacement field $\bm{u}$ or the stress field $\bm{P}$. Upright bold symbols are used for vectors or tensors representing algebraic or discretized quantities, such as the identity matrix $\mathbf{I}$, discrete stress field snapshots $\mathbf{P}$ or the snapshot tensor $\mathbf{S}$. When $\mathbf{A}$ is a $N$th-order tensor with indices $x_1, x_2, ..., x_N$, then $\mathbf{A}_{x_1, x_2, ..., x_n}$ with $n\leq N$ denotes the same tensor with the first $n$ indices held fixed and is of ($N-n$)th-order. As an example, when $\mathbf{A}$ is a matrix, then $\mathbf{A}_i$ denotes the $i$th row vector and $A_{ij}$ the entry at the $i$th row and $j$th column. Macroscopic variables are denoted by an overline to distinguish them from microscopic quantities; otherwise, the notation described above is used for both micro- and macroscopic quantities.
\section{Homogenization Theory}
\label{sec:theory}

\subsection{Macroscopic simulation}
Consider a solid body $\bar{\Omega}$ that is deformed under prescribed boundary conditions. Under the deformation, each point $\bar{\bm{X}}$ of the undeformed body is continuously mapped onto the coordinates $\bar{\bm{x}}$ of the deformed body. The macroscopic displacement for each point is defined as $\bar{\bm{u}}\coloneqq\bar{\bm{x}}-\bar{\bm{X}}$, and the macroscopic deformation gradient then follows as
\begin{align}
    \bar{\bm{F}} \coloneqq \deriv{\bar{\bm{x}}}{\bar{\bm{X}}} = \mathbf{I}+\deriv{\bar{\bm{u}}}{\bar{\bm{X}}}.
\end{align}
The governing partial differential equation (PDE) describing the deformation is given by the quasi-static linear momentum balance,
\begin{alignat}{2}
    \begin{aligned}
    \text{Div} \bar{\bm{P}} + \bar{\bm{B}} &= \bm{0}         &&\text{ on } \bar{\Omega}, \\
    \bar{\bm{P}}\bar{\bm{N}}      &= \bar{\bm{t}}_0 &&\text{ on } \partial\bar{\Omega}^N, \text{ and}\\
    \bar{\bm{u}}            &= \bar{\bm{u}}_0 &&\text{ on } \partial\bar{\Omega}^D,
    \end{aligned} \label{eq:macropde}
\end{alignat}
where $\bar{\bm{P}}$ is the macroscopic first Piola-Kirchhoff (PK1) stress tensor, $\bar{\bm{B}}$ are the body forces, $\bar{\bm{N}}$ is the outward normal on the surface of the body, $\bar{\bm{t}}_0$ and $\bar{\bm{u}}_0$ are the prescribed traction and displacement, and $\partial\bar{\Omega}^N$, $\partial\bar{\Omega}^D$ denote the Neumann and Dirichlet boundaries with $\partial\bar{\Omega} = \partial\bar{\Omega}^N \cup \partial\bar{\Omega}^D$ and $\partial\bar{\Omega}^N\cap\partial\bar{\Omega}^D=\emptyset$. The weak form of Eq.~\eqref{eq:macropde} is then

\begin{align}
    \bar{G} \coloneqq \int_{\bar{\Omega}} \deriv{\delta\bar{\bm{u}}}{\bar{\bm{X}}}:\bar{\bm{P}} dV - \int_{\bar{\Omega}} \bar{\bm{B}}\cdot \delta\bar{\bm{u}} dV - \int_{\partial\bar{\Omega}^N} \bar{\bm{t}}_0\cdot\delta\bar{\bm{u}} dA = 0, \qquad \forall \delta\bar{\bm{u}}\in H^1_0({\bar{\Omega}}),
\end{align}
where $\delta\bar{\bm{u}}\in H^1_0({\bar{\Omega}})=\{\bm{v} \in H^1(\bar{\Omega}) \ | \ \bm{v}=0 \text{ on } \partial\bar{\Omega}^D\}$ is a test function, and a solution for the displacement, $\bar{\bm{u}}\in H^1({\bar{\Omega}})$, is sought that fulfills $\bar{\bm{u}}=\bar{\bm{u}}_0$ on $\partial{{\bar{\Omega}}}^D$. To solve for the displacements, $\bar{\bm{u}}$, the linearization of $\bar{G}$ in the direction $\Delta\bar{\bm{u}}$ around the current deformation $\bar{\bm{u}}$ is needed,
\begin{align}
\begin{aligned}
    D\bar{G}\cdot \Delta \bar{\bm{u}} &= \int_{\bar{\Omega}} \deriv{(\delta\bar{\bm{u}})}{\bar{\bm{X}}}:\deriv{\bar{\bm{P}}}{\bar{\bm{F}}}:(D\bar{\bm{F}}\cdot\Delta\bar{\bm{u}}) dV \\
    &= \int_{\bar{\Omega}} \deriv{(\delta\bar{\bm{u}})}{\bar{\bm{X}}}:\bar{\bm{A}}:\deriv{\Delta \bar{\bm{u}}}{\bar{\bm{X}}} dV,
\end{aligned}
\end{align}
where $\bar{\bm{A}} \coloneqq \deriv{\bar{\bm{P}}}{\bar{\bm{F}}}$ is the macroscopic stiffness tensor. To model a certain material behavior, a constitutive relation which connects $\bar{\bm{P}}$ and $\bar{\bm{A}}$ to the deformation $\bar{\bm{F}}$ has to be provided.

\subsection{Microscopic simulation}
The microscopic simulation is defined on a representative volume element (RVE) $\Omega$. In first-order CH, it is assumed that the microscopic deformed points $\bm{x}$ are coupled to the macroscopic deformation by
\begin{align}
    \bm{x} \coloneqq \bar{\bm{F}}\bm{X} + \bm{w}(\bm{X}), \label{eq:microscopiccurrentpos}
\end{align}
where $\bm{w}(\bm{X})$ is a zero-mean microscopic fluctuation displacement field. Note that the macroscopic deformation gradient $\bar{\bm{F}}$ depends on the macroscopic coordinates $\bar{\bm{X}}$. For conciseness, this dependence is ommited here and the following equations are given for a fixed $\bar{\bm{X}}$. The microscopic displacement is then given as
\begin{align}
    \bm{u} \coloneqq (\bar{\bm{F}}-\mathbf{I})\bm{X} + \bm{w}(\bm{X}),\label{eq:displacementcoupling}
\end{align}
and the microscopic deformation gradient as
\begin{align}
    \bm{F} \coloneqq \deriv{\bm{x}}{\bm{X}} = \bar{\bm{F}} + \deriv{\bm{w}}{\bm{X}} \label{eq:deformationcoupling}.
\end{align}
Analogous to the above, the microscopic deformation is governed by the quasi-static linear momentum balance,
\begin{alignat}{2}
\begin{aligned}
    \text{Div} \bm{P}      &= \bm{0}         &&\text{ on } \Omega, \\
    \bm{P}\bm{N}           &= \bm{t}_0 &&\text{ on } \partial\Omega^N, \text{ and} \\
    \bm{u}                  &= \bm{u}_0 &&\text{ on } \partial\Omega^D,
\end{aligned} \label{eq:micropde}
\end{alignat}
where body forces are neglected for simplicity. The weak form is given as
\begin{align}
    G = \int_\Omega \deriv{\delta\bm{u}}{\bm{X}}:\bm{P} dV - \int_{\partial\Omega^N} \bm{t}_0\cdot\delta\bm{u} dA = 0,\qquad \forall \delta\bm{u}\in H^1_0(\Omega),
\end{align}
and the linearization of $G$ in the direction $\Delta\bm{u}$ around the current state $\bm{u}$ is
\begin{align}
\begin{aligned}
    DG\cdot \Delta \bm{u} &= \int_\Omega \deriv{(\delta\bm{u})}{\bm{X}}:\deriv{\bm{P}}{\bm{F}}:(D\bm{F}\cdot\Delta\bm{u}) dV \\
    &= \int_\Omega \deriv{(\delta\bm{u})}{\bm{X}}:\bm{A}:\deriv{\Delta \bm{u}}{\bm{X}} dV.
\end{aligned}
\end{align}

In contrast to the macroscale, the materials are fully specified by a constitutive law in the microscale. In this work, we consider a hyperelastic Neo-Hookean material model with strain energy density function
\begin{align}
    W(\bm{F}, \bm{\mu}) &= C_1(\text{Tr}(\bm{C})-3-2\ln{J})+D_1(J-1)^2, \label{eq:neohookeanenergy}
\end{align}
where $C_1$ and $D_1$ are the material parameters stored in $\bm{\mu}=[C_1,D_1]^T$, $\text{Tr}(\bullet)$ denotes the trace of a tensor, $\bm{C}=\bm{F}^T\bm{F}$ the right Cauchy-Green deformation tensor and $J=\text{det}(\bm{F})$ is the determinant of $\bm{F}$ respectively. The material parameters $C_1$ and $D_1$ are related to the Young's modulus $E$ and Poisson ratio $\nu$ by
\begin{align}
    E = \frac{2C_1 (3D_1 + 2C_1)}{C_1+D_1}, \quad \nu=\frac{D_1}{2(C_1+D_1)}. \label{eq:lameconstants}
\end{align}
The PK1 stress and stiffness tensor can then be derived from Eq.~\eqref{eq:neohookeanenergy},
\begin{align}
    \bm{P} = \deriv{W}{\bm{F}}, \quad \bm{A} = \deriv{\bm{P}}{\bm{F}} = \deriv[2]{W}{\bm{F}}.
\end{align}

\subsection{Scale coupling}
\begin{figure}[t]
    \centering
    \includegraphics[width=0.5\textwidth]{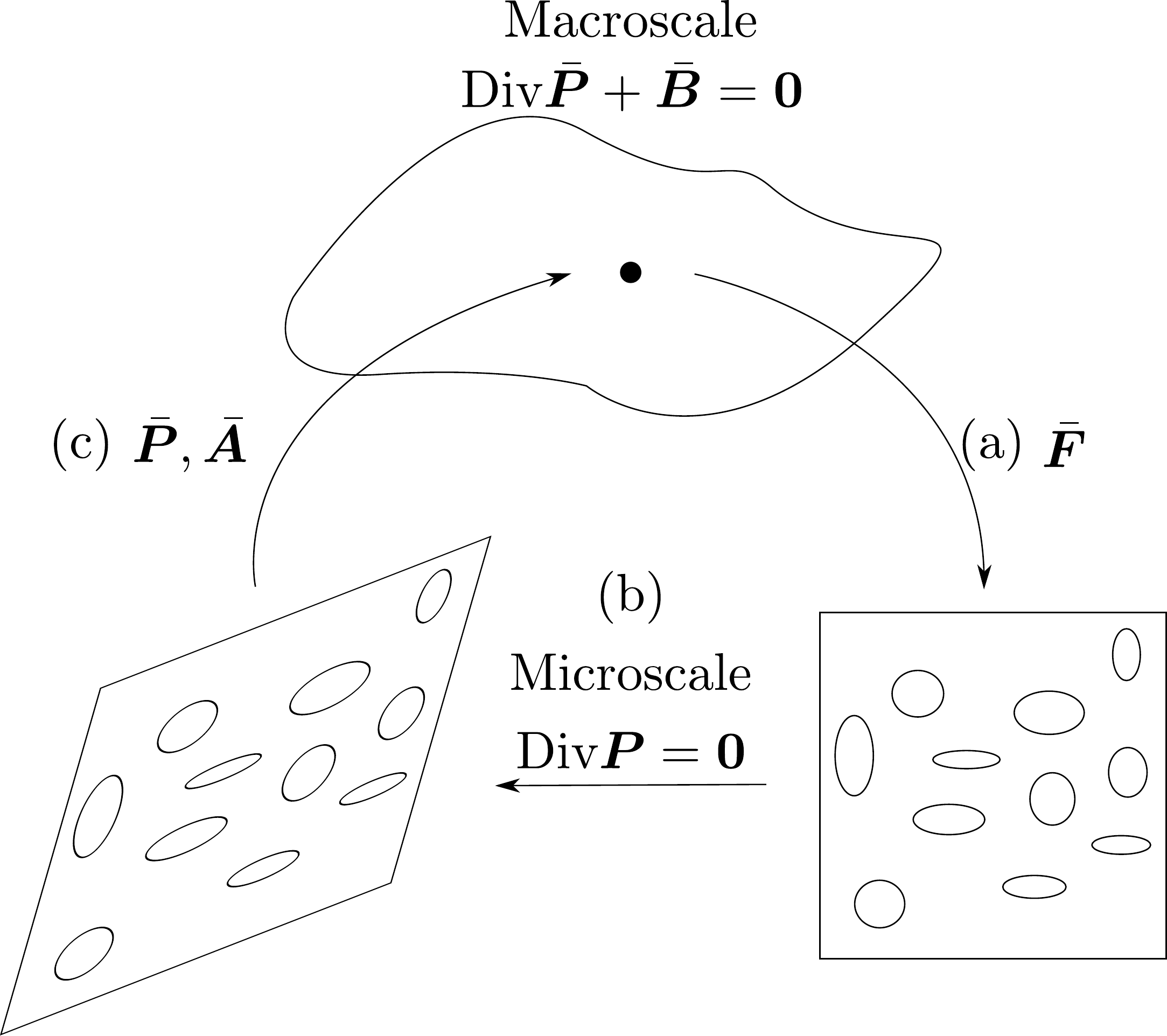}
    \caption{Scale coupling. (a) The macroscale simulation transfers a macroscopic deformation gradient to the microscale. (b) The microscopic response is subsequently computed and (c) the effective stress and stiffness are transferred back to the macroscale. The macroscopic solver does not require the microscopic deformation, hence it is beneficial to directly construct a surrogate model for the microscopic stress.}
    \label{fig:micromacro}
\end{figure}
The aim of the microscopic simulation is to replace the macroscopic constitutive model $\bar{\bm{P}}(\bar{\bm{X}}, \bar{\bm{F}})$. For every quadrature point in the macrostructure, the macroscopic simulation transfers the macroscopic deformation gradient to the microscopic simulation, which in turn returns an effective stress and stiffness tensor; see Fig.~\ref{fig:micromacro} for a visualization. The displacements of both scales are coupled according to Eq.~\eqref{eq:displacementcoupling}. Furthermore, the \textbf{Hill-Mandel Condition},
\begin{align}
    \left< \bm{P}:\delta \bm{F}\right> = \bar{\bm{P}}:\delta \bar{\bm{F}} \label{eq:hillmandel},
\end{align}
states that the virtual work exerted on both scales has to be the same; here, $\left<\{\bullet\}\right>=|\Omega|^{-1}\int_\Omega \{\bullet\} dV$ denotes the volumetric averaging operator with $|\Omega|$ the volume of the RVE $\Omega$.

It can be shown that the condition is always fulfilled by introducing appropriate boundary conditions for the microscopic problem. One set of boundary conditions that fulfill Eq.~\eqref{eq:hillmandel} are linear displacement boundary conditions, where the fluctuation displacement $\bm{w}$ on the RVE boundary is assumed to be zero. With Eq.~\eqref{eq:microscopiccurrentpos}, the following Dirichlet boundary condition follows:
\begin{align}
    \bm{u} = \bm{x} - \bm{X} = (\bar{\bm{F}}-\mathbf{I})\bm{X} \quad \text{ on } \partial\Omega. \label{eq:dbc}
\end{align}

In \cite{Saeb2016}, it is shown that, by prescribing Eq.~\eqref{eq:dbc} on the boundaries, the macroscopic deformation gradient is always equal to the averaged microscopic deformation gradient, i.e. $\bar{\bm{F}}=\left<\bm{F}\right>$. It then follows that the averaged microscopic stress has to be equal to the macroscopic stress:
\begin{align}
    \left<\bm{P}\right> = \left<\deriv{W}{\bm{F}}\right>=\deriv{\bar{W}}{\bar{\bm{F}}}:\deriv{\bar{\bm{F}}}{\bm{F}}=\deriv{\bar{W}}{\bar{\bm{F}}}=\bar{\bm{P}}.
\end{align}
However, the averaged stiffness tensor is generally not equal to the macroscopic stiffness tensor due to the changes of $\bm{w}$ inside $\Omega$, i.e.,
\begin{align}
    \left<\bm{A}\right> \not = \bar{\bm{A}},
\end{align}
thus complicating the calculation of the correct effective stiffness \cite{Saeb2016}. A popular way to compute the stiffness with a perturbation method is suggested in \cite{Temizer2008}, which essentially approximates the stiffness with a finite difference scheme.

\section{Reduced order model}
\label{sec:reducedbasis}
The two-scale simulation presented above can be solved using the finite element method (FEM) on both scales, leading to a nested FE scheme, also known as FE$^2$. Although effective, FE$^2$ method tends to be computationally very expensive. To lower the computational cost, we approximate the microscopic problem using a non-intrusive reduced order model, constructed by a proper orthogonal decomposition (POD) and a Gaussian process regression (GPR), which we will refer to as P-PODGPR later. The proposed method differs from existing works in the literature (see e.g. \cite{Guo2018, Kast2020, Chakir2018, Swischuk2019}) in that the non-intrusive reduced order model is constructed directly for the stresses instead of the displacements or strains. As seen in Fig.~\ref{fig:micromacro}, both displacement and strain are not required by the macroscopic solver and, moreover, the stresses still need to be calculated, which means that the microscopic constitutive law has to be implemented anew, negating the advantages of a non-intrusive method. Moreover, depending on the material model the evaluation might be costly and compromise the speed up significantly. It is also not obvious how one would derive the effective stiffness for that case. On the other hand, reducing the stresses directly circumvents the above-mentioned disadvantages, and both the effective stress and stiffness can be rapidly obtained.

\subsection{Proper orthogonal decomposition (POD)}
Principal component analysis (PCA) is a powerful tool in data science to find a low-dimensional approximation in Euclidean space. By introducing an appropriate function metric, PCA can be generalized for functions in Hilbert spaces and is referred to as proper orthogonal decomposition (POD). POD can utilize the correlation of solutions of a problem for different parameters to find a low-dimensional orthonormal basis, thus reducing the number of unknowns to a very small fraction of the dimension of the original high-fidelity model. The procedure for computing the low-dimensional basis of the microscopic stress $\bm{P}\in L^2(\Omega)$ is outlined in the following.

\subsubsection{Basis construction}
After $N_{\rm{pod}}$ high-fidelity simulations for different parameters have been carried out, the microscopic stress in all quadrature points $\mathbf{P} \in \mathbb{R}^{N_{\rm{qp}}\times3\times3}$, where $N_{\rm{qp}}$ is the total number of quadrature points in the high-fidelity model, is collected for each solution in the snapshot tensor $\mathbf{S}\in \mathbb{R}^{N_{\rm{qp}}\times3\times3\times N_{\rm{pod}}}$,
\begin{align}
    \mathbf{S} = \begin{bmatrix} \mathbf{P}^{(1)}, \mathbf{P}^{(2)}, \dots, \mathbf{P}^{(N_{\rm{pod}})}\end{bmatrix}.
\end{align}

Next, the correlation matrix $\mathbf{C}\in \mathbb{R}^{N_{\rm{pod}}\times N_{\rm{pod}}}$ can be formed by computing the $L^2$-inner product between every pair of two snapshots,
\begin{align}
\begin{aligned}
    \text{C}_{ij} &= \left(\bm{P}^{(i)}, \bm{P}^{(j)}\right)_{L^2(\Omega)} \\
    &= \int_\Omega \bm{P}^{(i)}:\bm{P}^{(j)} dV \\
    &= \sum_{q=1}^{N_{\rm{qp}}} w_q \mathbf{P}^{(i)}_q : \mathbf{P}^{(j)}_q,
\end{aligned}
\end{align}
where $(\bullet, \bullet)_{L^2(\Omega)}$ denotes the $L^2$-inner product and $i,j=1,2,\dots,N_{\rm{pod}}$. The weights $w_q$ are the integration weights corresponding to each quadrature point; they only depend on the mesh and can be easily computed. Note the subtle difference between italic and nonitalic characters used, corresponding to $L^2(\Omega)$ functions and their discrete counterparts respectively.

After obtaining the correlation matrix, its eigenvalues $\lambda_l$ and eigenvectors $\mathbf{v}_l$, $l=1, 2,\dots, \min{(L,N_{\rm{pod}})}$ are computed, where $L$ is the maximum number of basis functions specified by the user. The number of basis functions is often chosen from the criterion
\begin{align}
    \frac{\sum_{l=1}^L \lambda_l}{\sum_{l=1}^{N_{\rm{pod}}} \lambda_l}>\mathcal{E}_{\rm{pod}},
\end{align}
where $\mathcal{E}_{\rm{pod}}$ corresponds to the energy captured by $L$ basis functions and is specified by the user. Moreover, the eigenvalues can reveal if a problem is reducible using POD: if the eigenvalues decay rapidly, the solution space can be accurately captured by a few basis functions.
Finally, the $l$th POD basis function $\mathbf{B}_l \in \mathbb{R}^{N_{\rm{qp}}\times3\times3} $ is found with
\begin{align}
    \mathbf{B}_l &= \frac{1}{\sqrt{\lambda_l}} \mathbf{S} \mathbf{v}_l.
\end{align}
By construction, the POD basis is orthonormal, i.e.
\begin{align}
    (\bm{B}_i, \bm{B}_j)_{L^2(\Omega)} = \begin{cases} 1 & i=j \\ 0 & i\not=j
    \end{cases}. \label{eq:l2orthogonality}
\end{align}

\subsubsection{Approximation of stress and stiffness}
With the POD basis functions, $\mathbf{P}$ can be approximated with
\begin{align}
    \mathbf{P}^\text{RB}(\bar{\bm{F}}, \bm{\mu}) = \sum_{l=1}^L \alpha_l(\bar{\bm{F}},\bm{\mu}) \mathbf{B}_l, \label{eq:stressapprox}
\end{align}
where the loading $\bar{\bm{F}}$ and material parameters $\bm{\mu}$ are listed separately, as the partial derivative over the loading parameters yields the stiffness tensor, whereas the partial derivative with respect to $\bm{\mu}$ could be used for optimization or inverse problems. Since the basis functions $\mathbf{B}_l$ are constant, only the coefficients $\alpha_l\in \mathbb{R}$ depend on the parameters and need to be found for a new parameter value. Note that with this approach the ability to determine the microscopic displacement, energy and stiffness tensor is lost, if the constitutive relation $\bm{P}(\bm{F})$ can not be inverted. However, if knowledge of the deformation is required, a similar procedure using displacement or deformation gradient snapshots could be implemented \cite{Guo2018, Kast2020}. This should in general work well if the stresses can be approximated accurately, because the stress is a non-linear function of the displacement and therefore should be more difficult to approximate.

By averaging the microscopic stress, the macroscopic stress $\bar{\bm{P}}\in \mathbb{R}^{3\times 3}$ is found:
\begin{align}
    \begin{aligned}
        \bar{\bm{P}} &= \left< \bm{P}^\text{RB} \right> \\
        &= |\Omega|^{-1}\int_\Omega \bm{P}^\text{RB} dV\\
        &= |\Omega|^{-1}\sum_{q=1}^{N_{\rm{qp}}} w_q \mathbf{P}^\text{RB}_q\\
        &= |\Omega|^{-1}\sum_{q=1}^{N_{\rm{qp}}} w_q \left(\sum_{l=1}^L \alpha_l(\bar{\bm{F}},\bm{\mu}) \mathbf{B}_{lq}\right).
    \end{aligned}
\end{align}
Lastly, the effective stiffness $\bar{\bm{A}}\in \mathbb{R}^{3\times3\times3\times3}$ is determined by differentiating the macroscopic stress over the macroscopic deformation gradient,
\begin{align}
\begin{aligned}
    \bar{\bm{A}} &= \deriv{\bar{\bm{P}}}{\bar{\bm{F}}} \\
    &= |\Omega|^{-1}\deriv{\left(\sum_{q=1}^{N_{\rm{qp}}} w_q \left(\sum_{l=1}^L \alpha_l(\bar{\bm{F}},\bm{\mu}) \mathbf{B}_{lq}\right)\right)}{\bar{\bm{F}}} \\
    &= |\Omega|^{-1}\sum_{q=1}^{N_{\rm{qp}}} w_q \left(\sum_{l=1}^L \mathbf{B}_{lq}\otimes \deriv{\alpha_l(\bar{\bm{F}},\bm{\mu})}{\bar{\bm{F}}} \right). \label{eq:effectivestiffness}
\end{aligned}
\end{align}

As seen above, by introducing the approximation Eq.~\eqref{eq:stressapprox}, the macroscopic stiffness tensor can be naturally derived when the derivative of the coefficients $\alpha_l$ with respect to $\bar{\bm{F}}$ is available. Moreover, the derivatives of stress with respect to the material parameters $\bm{\mu}$ can be obtained analogously and used in optimization or inverse problems.

\paragraph{Remark}
As this surrogate model is used to replace the microstructural simulation, it is sufficient to construct the basis for macroscopic stretches $\bar{\bm{U}}$ instead of $\bar{\bm{F}}$, since the macroscopic deformation gradient can be split into $\bar{\bm{F}}=\bar{\bm{R}}\bar{\bm{U}}$ by using the polar decomposition, where $\bar{\bm{R}}$ accounts for rotations. From the principle of material objectivity it follows,
\begin{align}
    \bar{\bm{P}}(\bar{\bm{F}})&=\bar{\bm{R}}\bar{\bm{P}}(\bar{\bm{U}})\\
    \bar{A}_{ijkl}(\bar{\bm{F}})&=\sum_{m=1}^3\sum_{n=1}^3 \bar{R}_{im}\bar{A}_{mjnl}(\bar{\bm{U}})\bar{R}_{kn},
\end{align}
see \cite{Kunc2019}. Due to the symmetry of $\bar{\bm{U}}$, the number of parameters reduces from 4 to 3 in 2D or from 9 to 6 in 3D. In the following, $\bar{\bm{F}}$ is therefore replaced by $\bar{\bm{U}}$ and all snapshots are assumed to have been obtained for different $\bar{\bm{U}}$ and $\bm{\mu}$.

\subsection{Regression model}
The mapping $\alpha_l(\bar{\bm{U}},\bm{\mu})$ for all parameters can be constructed by means of a regression, using the training data that have been collected for POD. If the regression quality is insufficient, more data has to be generated.

\subsubsection{Data preparation}
The stress data $\mathbf{P}^{(i)}$ collected was generated for parameters $(\bar{\bm{U}}^{(i)}, \bm{\mu}^{(i)})$ with $i=1,2,\dots,N_{\rm{reg}}$, where $N_{\rm{reg}}$ denotes the total number of snapshots available. The best approximation of the $i$th snapshot on the POD basis is given by
\begin{align}
    \mathbf{P}^{(i)} \approx \sum_{l=1}^L (\bm{P}^{(i)}, \bm{B}_l)_{L^2(\Omega)} \mathbf{B}_l,
\end{align}
and hence $\alpha_l^{(i)}=(\bm{P}^{(i)}, \bm{B}_l)_{L^2(\Omega)}$. All coefficients for all snapshots $\alpha_l^{(i)}$ are collected together with the parameter data; a mapping between the parameters and the POD coefficients is then constructed by regression, i.e. 
\begin{align}
    \hat{\alpha}_l: \mathcal{P} \rightarrow \mathbb{R}, (\bar{\bm{U}},\bm{\mu}) \mapsto \hat{\alpha}_l(\bar{\bm{U}},\bm{\mu}), \label{eq:alphamap}
\end{align}
where $\mathcal{P}$ denotes the parameter space. Due to the orthonormality of the POD basis functions, see Eq.~\eqref{eq:l2orthogonality}, the coefficients $\alpha_l$ are fully uncorrelated and a separate regression for each POD coefficient can be constructed independently. With this mapping, the stress $\mathbf{P}$ is approximated as
\begin{align}
    \hat{\mathbf{P}}^\text{RB}(\bar{\bm{U}}, \bm{\mu}) = \sum_{l=1}^L \hat{\alpha}_l(\bar{\bm{U}},\bm{\mu}) \mathbf{B}_l \label{eq:stressapproxGPR}.
\end{align}

\subsubsection{Gaussian process regression}
Since the material model considered here is hyperelastic and therefore history independent, each stress corresponds to exactly one set of deformations and therefore, a wide range of regression techniques can be used. Many different regression approaches have been used to approximate the coefficients, e.g. radial basis functions \cite{Guenot2013}, neural networks (NNs) \cite{Hesthaven2018}, and Gaussian process regression (GPR) \cite{Guo2018, Kast2020, Yang2020}. A systematic investigation on these three methods has been conducted in \cite{Berzins2020StandardizedProblems}. A comparison of different machine learning methods for regression was performed in \cite{Swischuk2019}. For this work, GPR is used since it offers some desirable properties:
\begin{enumerate}
    \item It can reconstruct the training data perfectly, i.e. it reproduces the exact solution at the training points.
    \item Depending on the chosen kernel, the regression function has a specified smoothness and its derivatives can be obtained analytically.
    \item The trained GPR model returns an uncertainty measure for each evaluation, which can be used to estimate the regression error or to develop an active learning scheme \cite{Liu2018a, Guo2018}.
\end{enumerate}

In GPR, a scalar regression function $f(\mathbf{X})$ with $\mathbf{X}\in\mathbb{R}^d$ is assumed to be distributed as a Gaussian process (GP) with a zero mean function and kernel $k_{\bm{\theta}}(\mathbf{X},\mathbf{X}')$,
\begin{equation}
f \sim \mathcal{GP}(0, k_{\bm{\theta}}(\mathbf{X},\mathbf{X}')).
\end{equation}
The form of the kernel $k_{\bm{\theta}}(\mathbf{X},\mathbf{X}')$ has to be chosen by the user, and each kernel has hyperparameters $\bm{\theta}$ that must be fitted to the data. In this work, we use, as in \cite{Guo2018}, the \textit{automatic relevance determination} (ARD) squared exponential kernel,
\begin{equation}
    k_{\bm{\theta}}(\mathbf{X},\mathbf{X}') = \sigma_f^2 \exp\left(-\frac{1}{2}\sum_{k=1}^d\frac{(X_k-X_k')^2}{l^2_k}\right),
\end{equation}
where $\bm{\theta}=[\sigma_f, l_1, l_2, \dots, l_d]$.

Given a finite number of data points $\{\mathbf{X}^{(i)},f(\mathbf{X}^{(i)})\}_{i=1}^{N_{\rm{reg}}}$, the optimal hyperparameters $\bm{\theta}$ can be determined with a maximum likelihood estimation \cite{Guo2018, Rasmussen2004}, hence yielding a prior GP $f$. By Bayesian inference, the posterior GP $f^*(\mathbf{x})$ can then be given as
\begin{align}
    f^*|(\mathbf{X},y) &\sim \mathcal{GP}(m^*, k^*), \\
    m^*(\mathbf{x}) &= k_{\bm{\theta}}(\mathbf{X}, \mathbf{x})^T k_{\bm{\theta}}(\mathbf{X}, \mathbf{X})^{-1} y(\mathbf{X}), \\
    k^*(\mathbf{x},\mathbf{x}') &= k_{\bm{\theta}}(\mathbf{x}, \mathbf{x}') - k_{\bm{\theta}}(\mathbf{X}, \mathbf{x})^T k_{\bm{\theta}}(\mathbf{X}, \mathbf{X})^{-1} k_{\bm{\theta}}(\mathbf{X}, \mathbf{x}'),
\end{align}
where $k_{\bm{\theta}}(\mathbf{X}, \mathbf{x}) = [k_{\bm{\theta}}(\mathbf{X}^{(1)}, \mathbf{x}), k_{\bm{\theta}}(\mathbf{X}^{(2)}, \mathbf{x}), \dots, k_{\bm{\theta}}(\mathbf{X}^{(N_{\rm{reg}})}, \mathbf{x})]^T$, $k_{\bm{\theta}}(\mathbf{X}, \mathbf{X}) = [k_{\bm{\theta}}(\mathbf{X}^{(i)}, \mathbf{X}^{(j)})]^{N_{\rm{reg}}}_{i,j=1}$, and $y(\mathbf{X})=[f(\mathbf{X}^{(1)}), f(\mathbf{X}^{(2)}), \dots, f(\mathbf{X}^{(N_{\rm{reg}})})]^T$. For any arbitrary input $\mathbf{x}$, a distribution $\mathcal{N}(m^*(\mathbf{x}),k^*(\mathbf{x},\mathbf{x}))$ can thus be obtained.

A GPR is separately performed for each POD basis coefficient, yielding in total $L$ GPR models. The ARD kernel is infinitely smooth, hence the Jacobian of $\alpha_l$ can be easily obtained and Eq.~\eqref{eq:effectivestiffness} is fully specified. With the trained GPR models and the POD basis, the surrogate model for the microscopic simulation is complete. Given a parameter, the microscopic stress field can be rapidly evaluated, from which the effective stress and its derivative can be derived. Therefore the surrogate model can be considered a learned constitutive model, where additionally the stress field on the microscale is still accessible.

As indicated earlier, the non-intrusive reduced order model will be referred to as P-PODGPR.

\subsubsection{Error estimate}
For brevity of notation, we use the symbol $\bm{\pi}$ to denote both the loading and material parameters in this section. Using the Cauchy-Schwarz inequality, it can be shown that the total error between the high fidelity simulation and the approximation by P-PODGPR for a parameter $\bm{\pi}\in\mathcal{P}$ consists of the projection error and regression error,
\begin{align}
\begin{aligned}
    \epsilon_\text{P-PODGPR}(\bm{\pi}) 
    &= ||\bm{P}^\text{HF}(\bm{\pi})-\hat{\bm{P}}^\text{RB}(\bm{\pi})||_{L^2(\Omega)} \\
    &= ||\bm{P}^\text{HF}(\bm{\pi})-\bm{P}^\text{RB}(\bm{\pi})+\bm{P}^\text{RB}(\bm{\pi})-\hat{\bm{P}}^\text{RB}(\bm{\pi})||_{L^2(\Omega)} \\
    &\leq \underbrace{||\bm{P}^\text{HF}(\bm{\pi})-\bm{P}^\text{RB}(\bm{\pi})||_{L^2(\Omega)}}_\text{projection error}+\underbrace{||\bm{P}^\text{RB}(\bm{\pi})-\hat{\bm{P}}^\text{RB}(\bm{\pi})||_{L^2(\Omega)}}_\text{regression error}.
\end{aligned}
\end{align}
If the parameter space has been appropriately explored by the snapshots used for POD and the eigenvalues exhibit a rapid decay, the projection error can be reasonably expected to quickly approach zero for any parameter value as the number of basis functions $L$ is increased.

The squared regression error for the stresses can be rearranged,
\begin{align}
\begin{aligned}
    ||\bm{P}^\text{RB}(\bm{\pi})&-\hat{\bm{P}}^\text{RB}(\bm{\pi})||_{L^2(\Omega)}^2= \left\Vert\sum_{l=1}^L (\alpha_l-\hat{\alpha}_l) \bm{B}_l \right\Vert_{L^2(\Omega)}^2\\
    &= \left( \sum_{l=1}^L (\alpha_l-\hat{\alpha}_l) \bm{B}_l, \sum_{l'=1}^L (\alpha_{l'}-\hat{\alpha}_{l'}) \bm{B}_{l'}  \right)_{L^2(\Omega)} \\
    &= \sum_{l=1}^L \sum_{l'=1}^L (\alpha_l-\hat{\alpha}_l)(\alpha_{l'}-\hat{\alpha}_{l'}) \left(\bm{B}_l, \bm{B}_{l'}\right)_{L^2(\Omega)} \\
    &= \sum_{l=1}^L (\alpha_l-\hat{\alpha}_l)^2 \\
    &= ||\bm{\alpha}-\hat{\bm{\alpha}}||_2^2,
\end{aligned}\label{eq:regerror}
\end{align}
where the orthonormality of $\bm{B}_l$ was used in the second last line and $||\bullet||_2$ denotes the Euclidean norm. Hence, the regression error is simply equal to the Euclidean norm of the error of the coefficient vectors. This result is analogous to the result presented in \cite{Kast2020} for the displacement.

\subsection{Comparison to Neural Networks}
After the pioneering works by Ghaboussi et al.~\cite{Ghaboussi1991Knowledge-basedNetworks, Ghaboussi1998} and recent advances in Deep Learning, many papers have used methods of deep learning to extract a constitutive model from pairs of stress and deformation data by training a deep neural network (DNN), e.g. \cite{Huang2020, Le2015, Ghavamian2019, Mozaffar2019, Wu2020}. Frequently their data is collected from RVE simulations by applying boundary conditions that fulfill Eq.~\eqref{eq:hillmandel} and solving the microscopic simulation to generate training data, analogous to what is done in this work. However, one difference is that in this work the stress field is collected while the other works only collect the averaged stress and dispose of the stress field. A brief comparison between both approaches is given below:
\begin{enumerate}
    \item \textbf{Training}: The DNN approach essentially performs a regression on the stress and deformation data. This means that a mapping from $\mathbb{R}^{3\times3}\rightarrow\mathbb{R}^{3\times3}$ (or $\mathbb{R}^6\rightarrow\mathbb{R}^6$ in the case where the stress and deformation measures are symmetric) has to be learned. P-PODGPR, on the other hand, uses the correlation of the microscopic stress field solutions and finds a few uncorrelated coefficients which can be independently learned, i.e. a mapping from $\mathbb{R}^{3\times3}\rightarrow\mathbb{R}$ or $\mathbb{R}^6\rightarrow\mathbb{R}$ is learned.
    \item \textbf{Implementation}: Both surrogate models, after they have been trained, can be easily adopted into any simulation software as they are both non-intrusive and therefore entirely independent of the software used to solve the microscale simulation.
    \item \textbf{Evaluation}: The DNN approach needs to compute one forward pass through the neural network to get the effective stress for a given deformation. The effective stiffness can then be computed with one backward pass with automatic differentiation. However, the microstructural stresses, which are desirable for finding local stress concentrations and designing improved microstructures, cannot be obtained. In the P-PODGPR method the stress field is fully obtained by evaluating $L$ GPR models, from which the effective stress can be computed. The effective stiffness is obtained with Eq.~\eqref{eq:effectivestiffness} as the GPR model also supports an analytical way of computing the derivative of $\alpha_l$ over the deformation.
\end{enumerate}
\section{Results}
\label{sec:results}


To demonstrate the performance of P-PODGPR and the influence of the number of basis functions $L$, number of samples used for the basis construction $N_{\rm{pod}}$, and number of samples used for the regression $N_{\rm{reg}}$, two single-scale examples with different 2D RVEs are presented. The third example involves a two-scale problem, in which the results obtained with P-PODGPR are compared with the FE$^2$ solution.

\subsection{Single-scale RVE simulation}
To measure the accuracy of P-PODGPR on test data, the following relative error measure for the effective stress is defined:
\begin{align}
    \epsilon_{\bar{\mathbf{P}}} \coloneqq \frac{\left\Vert\bar{\mathbf{P}}^\text{HF}-\bar{\hat{\mathbf{P}}}^\text{RB}\right\Vert_{\rm{F}}}{\left\Vert\bar{\mathbf{P}}^\text{HF}\right\Vert_{\rm{F}}},
\end{align}
where $\left\Vert\bullet\right\Vert_{\rm{F}}$ denotes the Frobenius norm, $\bar{\mathbf{P}}^\text{HF}$ the effective stress from the high-fidelity simulation and $\bar{\hat{\mathbf{P}}}^\text{RB}$ the effective stress resulting from P-PODGPR. The projection error of the basis is attained when $\bar{\hat{\mathbf{P}}}^\text{RB}=\bar{\mathbf{P}}^\text{RB}$, where $\bar{\mathbf{P}}^\text{RB}$ is the effective stress after projecting the high fidelity solution onto the POD basis when the solution is known. As the regression error Eq.~\eqref{eq:regerror} goes towards zero, $\bar{\hat{\mathbf{P}}}^\text{RB}$ tends towards $\bar{\mathbf{P}}^\text{RB}$, so the projection error can be interpreted as a lower bound of the solution from P-PODGPR. All microscopic simulations were conducted with the FE framework MOOSE \cite{permann2020moose}.

\subsubsection{Porous material}
\label{subsubsec:porous}

\begin{figure}[tb]
     \centering
     \begin{subfigure}[b]{0.4\textwidth}
         \centering
         \includegraphics[width=\textwidth]{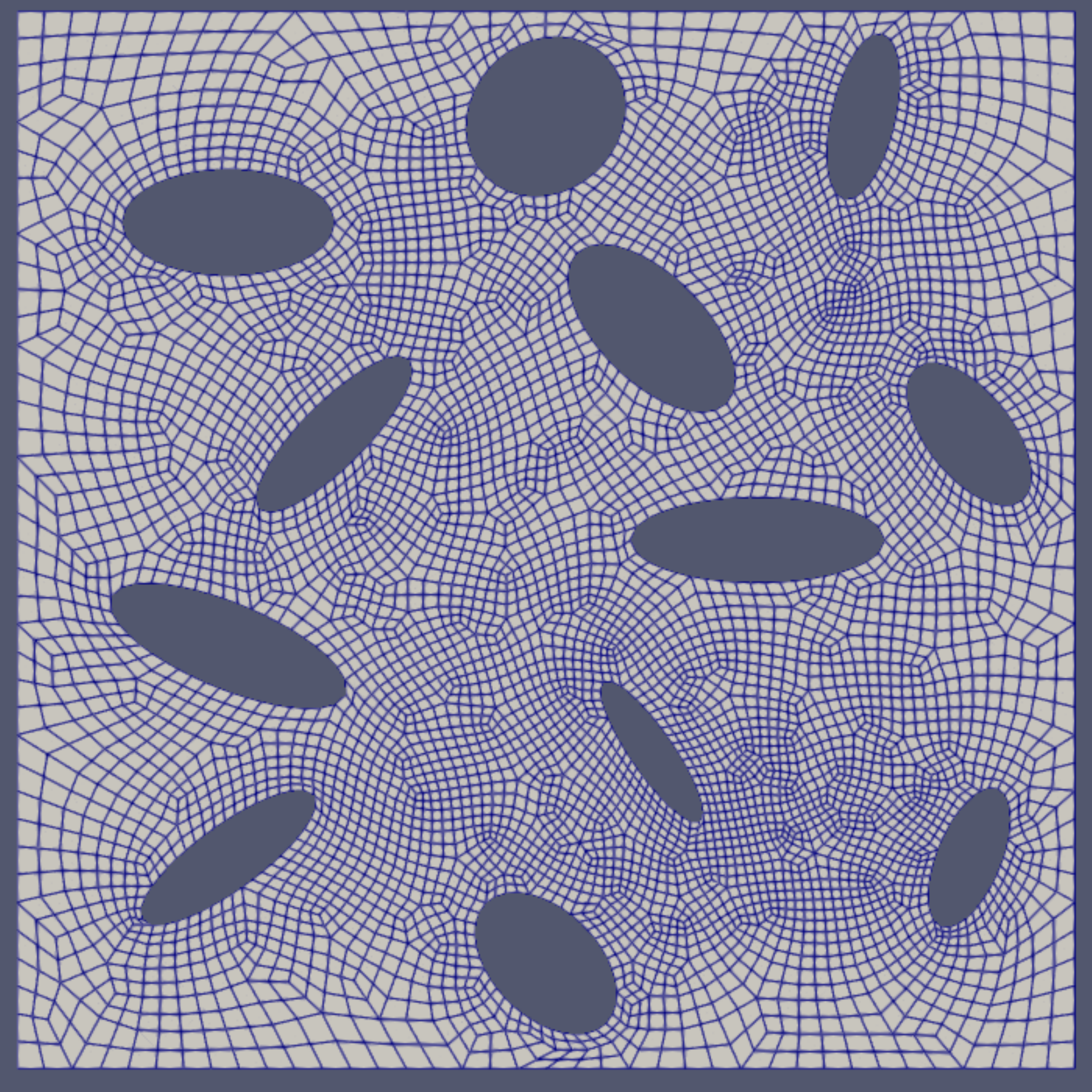}
         \caption{}
         \label{fig:rve1mesh}
     \end{subfigure}
     \hfill
     \begin{subfigure}[b]{0.535\textwidth}
         \centering
         \includegraphics[width=\textwidth]{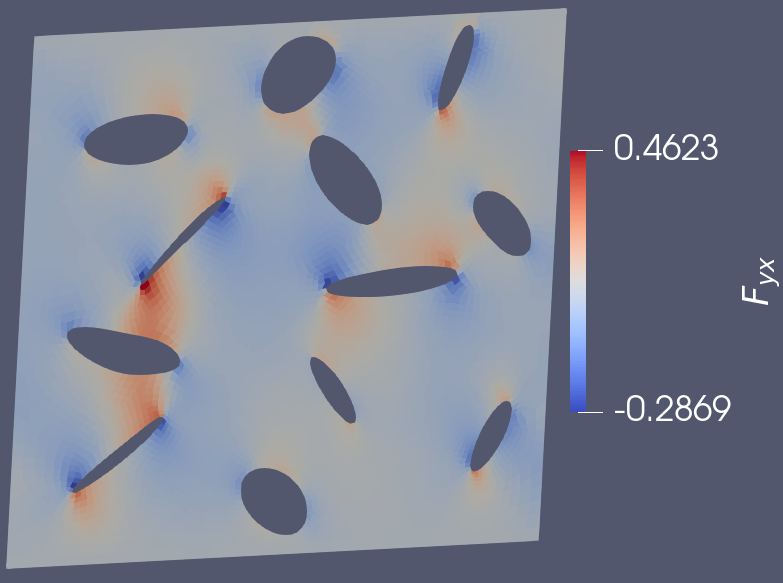}
         \caption{}
         \label{fig:examplePore}
     \end{subfigure}
    \caption{Porous material: a) The considered microstructure consists of a Neo-Hookean matrix material with a volume ratio of 86\%. The mesh consists of 6000 bilinear elements and 6293 nodes. b) An exemplary deformation with $[\bar{U}_{xx}, \bar{U}_{yy},\bar{U}_{xy}]=[0.95, 0.95, 0.05]$. The $yx$-component of the local deformation gradient ranges from $[-0.2869, 0.4623]$ and is hence much larger than the prescribed boundary deformation.}
\end{figure}


A porous microstructure as depicted in Fig.~\ref{fig:rve1mesh} is considered for the first example, where the pores account for 14\% of the total area. Four-node quadrilateral elements with four quadrature points are employed. The matrix material is modeled as a Neo-Hookean material with material parameters $C_1=1$ and $D_1=1$. Linear displacement boundaries with $\bar{\bm{U}}-\mathbf{I}\in [-0.05, 0.05]$ are considered. Due to the geometry of the problem, such a deformation leads to much higher local strains, see Fig.~\ref{fig:examplePore} for an exemplary deformation with $[\bar{U}_{xx}, \bar{U}_{yy},\bar{U}_{xy}]=[0.95, 0.95, 0.05]$. If larger displacements on the boundaries were considered, some pores might close, requiring contact detection. In this example, the material parameters $\bm{\mu}$ are fixed, hence this problem has three varying parameters.

\paragraph{Data generation}
To investigate the number of precomputations needed for an accurate representation, a set of 500 training snapshots for training P-PODGPR was sampled via a Sobol sequence sampling procedure and another set of 1000 test snapshots, to evaluate the accuracy of the surrogate model, was generated randomly from a uniform distribution. It was observed in \cite{Bessa2017} that Sobol sequence sampling fills the parameter space more evenly as compared to random sampling and thus provides better results.

\begin{figure}[tb]
    \centering
    \includegraphics[width=0.7\textwidth]{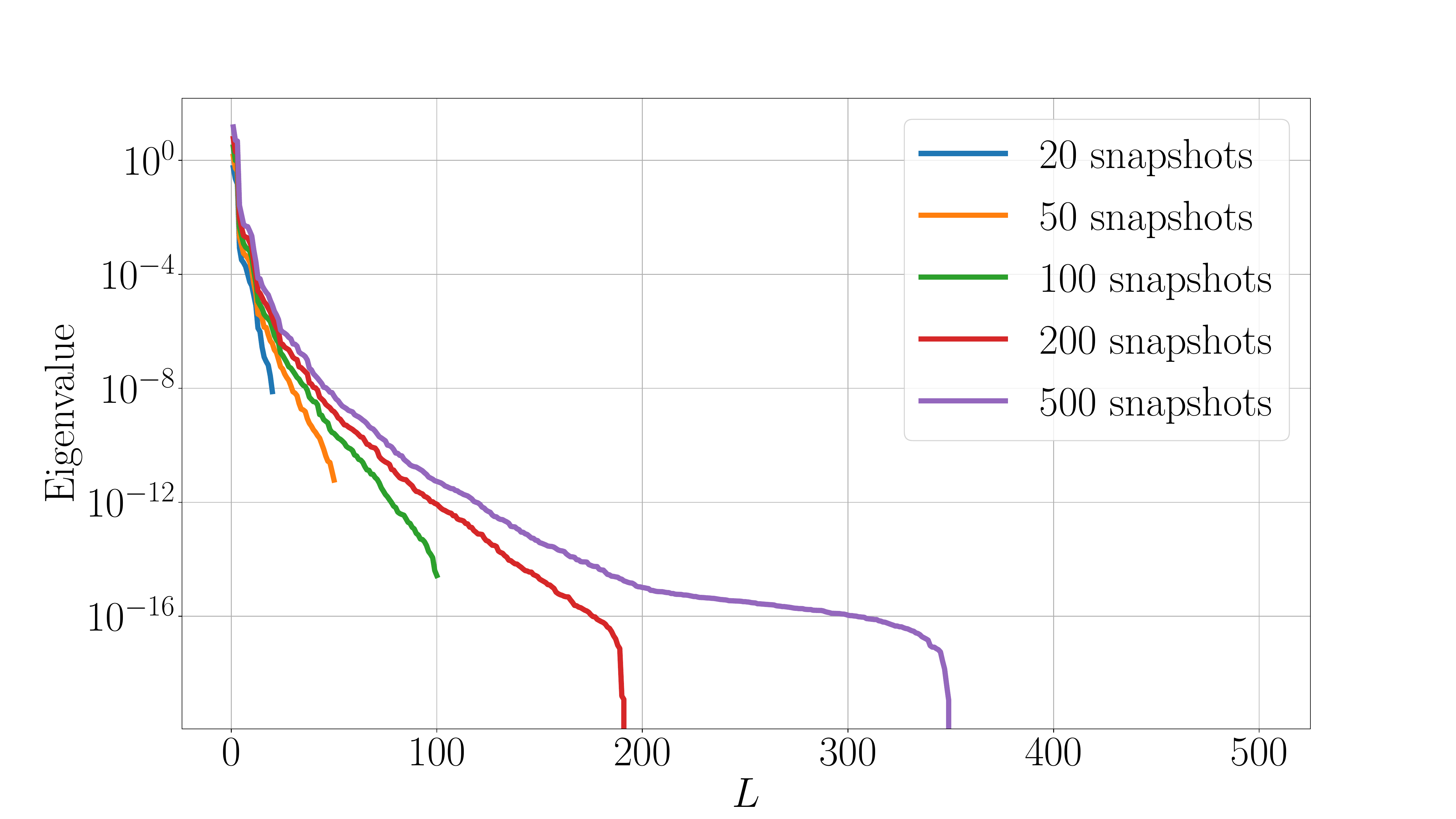}
    \caption{Porous material: Eigenvalues of the correlation matrix for different numbers of snapshots $N_{\rm{pod}}$ used for POD.}
    \label{fig:rve1eig}
\end{figure}

\begin{figure}[tb]
     \centering
     \begin{subfigure}[b]{0.3\textwidth}
         \centering
         \includegraphics[width=\textwidth]{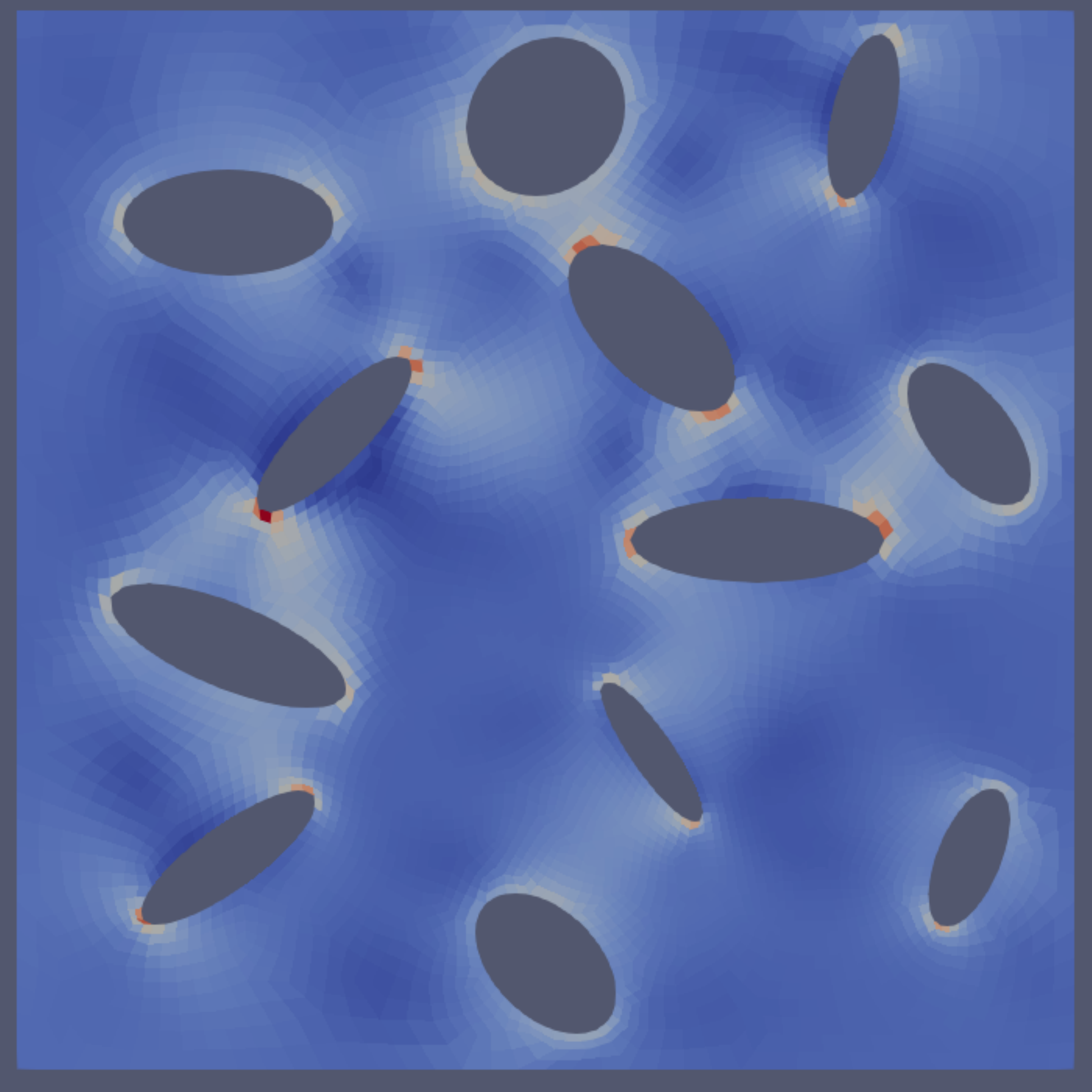}
         \caption{1st basis function}
         \label{fig:rve1basisfunc1}
     \end{subfigure}
     \hfill
     \begin{subfigure}[b]{0.3\textwidth}
         \centering
         \includegraphics[width=\textwidth]{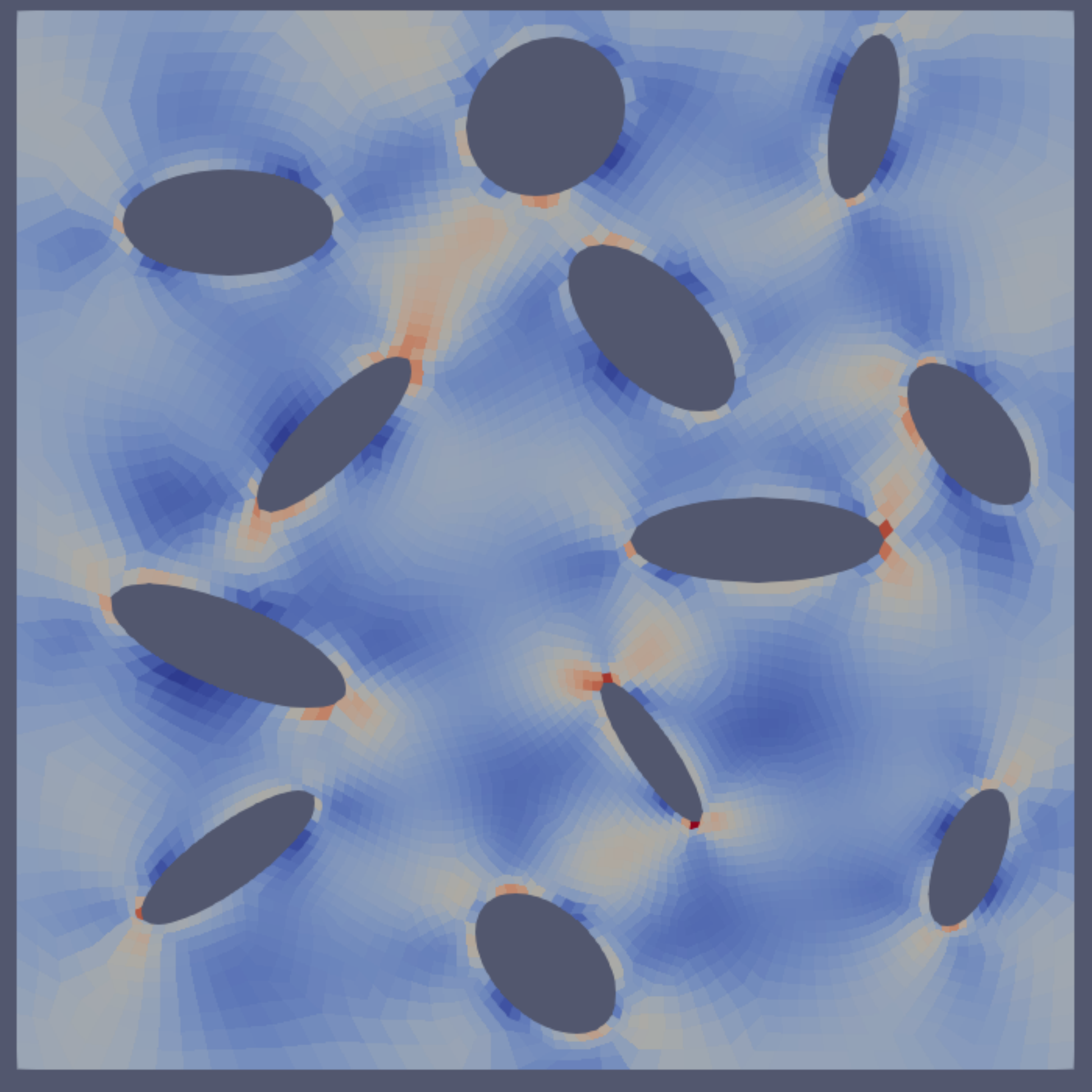}
         \caption{2nd basis function}
         \label{fig:rve1basisfunc2}
     \end{subfigure}
     \hfill
     \begin{subfigure}[b]{0.3\textwidth}
         \centering
         \includegraphics[width=\textwidth]{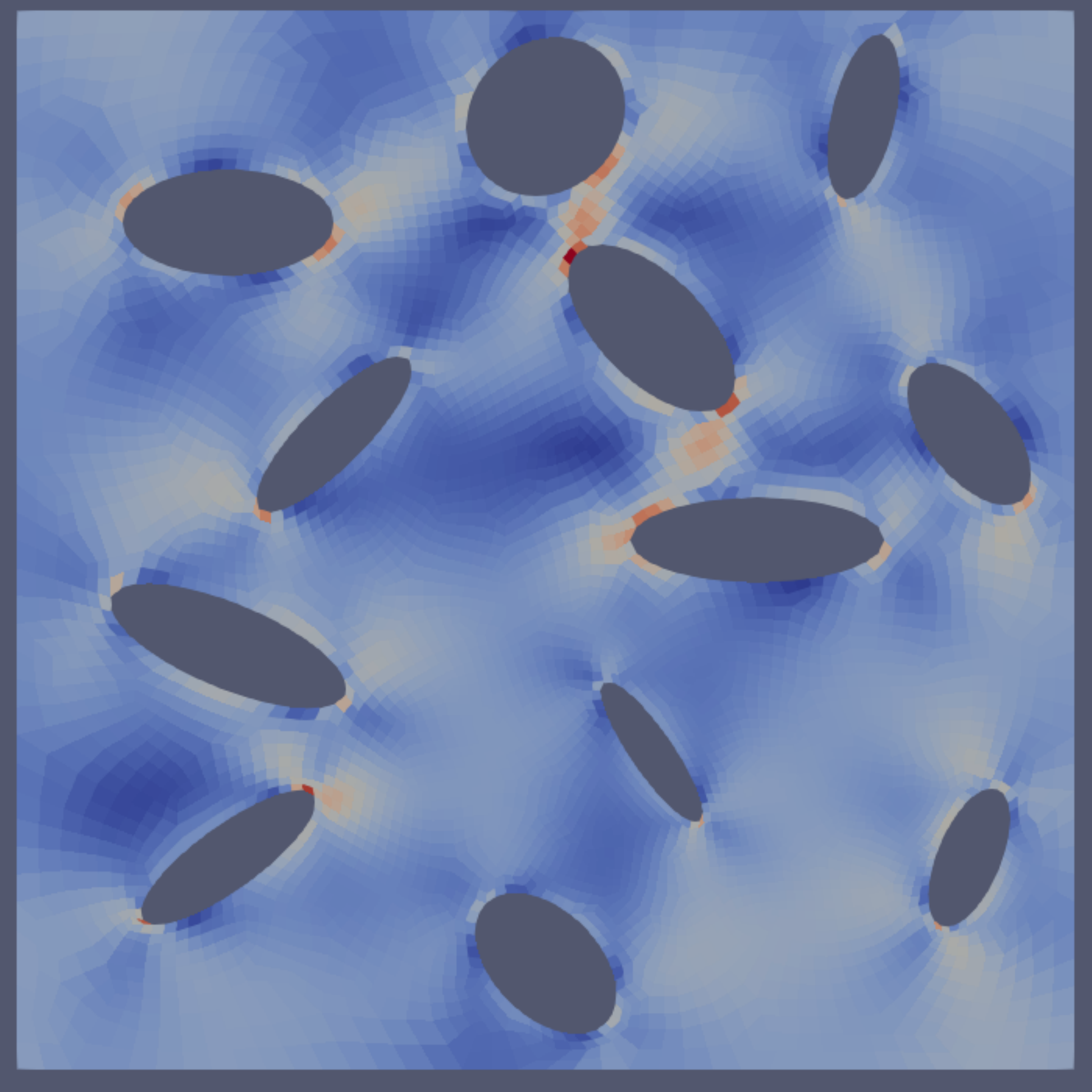}
         \caption{3rd basis function}
         \label{fig:rve1basisfunc3}
     \end{subfigure}
    \caption{Porous material: Von Mises stress of the first three POD basis functions of the microscopic stress field.}
    \label{fig:rve1basisfunc}
\end{figure}

\paragraph{Eigenvalues}
The eigenvalues of the correlation matrix for different numbers of training snapshots $N_{\rm{pod}}$ are plotted in Fig.~\ref{fig:rve1eig}. For all cases, an exponential decay is observed, indicating the reducibility of the problem. The von Mises stress of the first three POD basis functions is plotted in Fig.~\ref{fig:rve1basisfunc}. It can be seen that the basis functions specifically capture the stress concentrations around the pores. Subsequently, $L=20$ basis functions are considered which corresponds to an energy $\mathcal{E}_{\rm{pod}}$ of $99.9996\%$. 

\begin{figure}[p]
    \centering
    \includegraphics[width=0.95\textwidth]{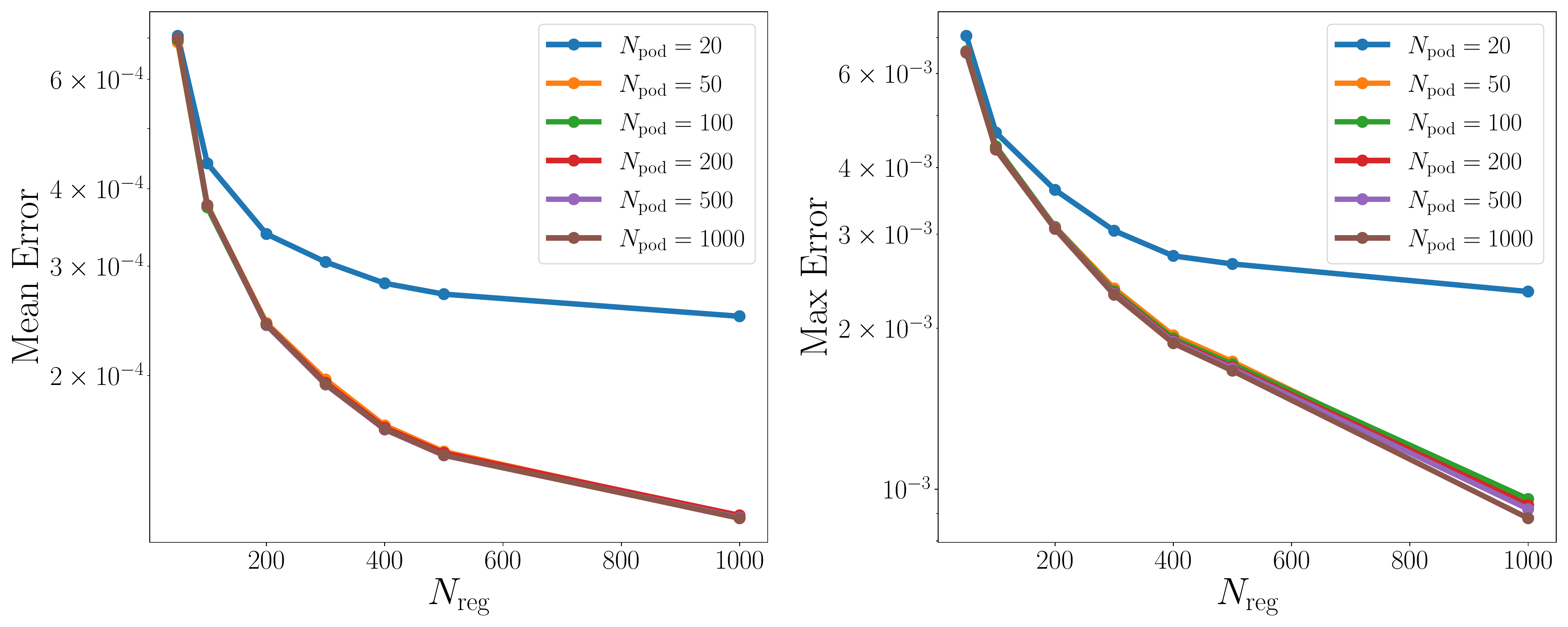}
    \caption{Porous material: Comparison of stress errors for different combinations of $N_{\rm{pod}}$ and $N_{\rm{reg}}$ for $L=20$.}
    \label{fig:rve1comparison}
\end{figure}

\paragraph{Influence of $N_{\rm{pod}}$ and $N_{\rm{reg}}$}
For $L=20$ basis functions, combinations of $N_{\rm{pod}}\in\{20,50,100,200,500\}$ and $N_{\rm{reg}}\in\{50,100,200,300,400,500\}$ were investigated. All 1000 test data were evaluated and the resulting mean and maximum error of the stresses are plotted in Fig.~\ref{fig:rve1comparison}. It can be observed that for all cases the mean error is below $0.1\%$. From $N_{\rm{pod}}=20$ to $N_{\rm{pod}}=50$, a significant improvement in the error can be observed. However, for $N_{\rm{pod}}>50$, the error only changes marginally. As $N_{\rm{reg}}$ is increased, more data becomes available for the regression and therefore the mapping in Eq.~\eqref{eq:alphamap} becomes increasingly more accurate. Nevertheless, even using only 50 snapshots for both basis and regression yields a mean error of 0.1\% and a maximum error of roughly 0.65\%.

\begin{figure}[p]
    \centering
    \includegraphics[width=0.95\textwidth]{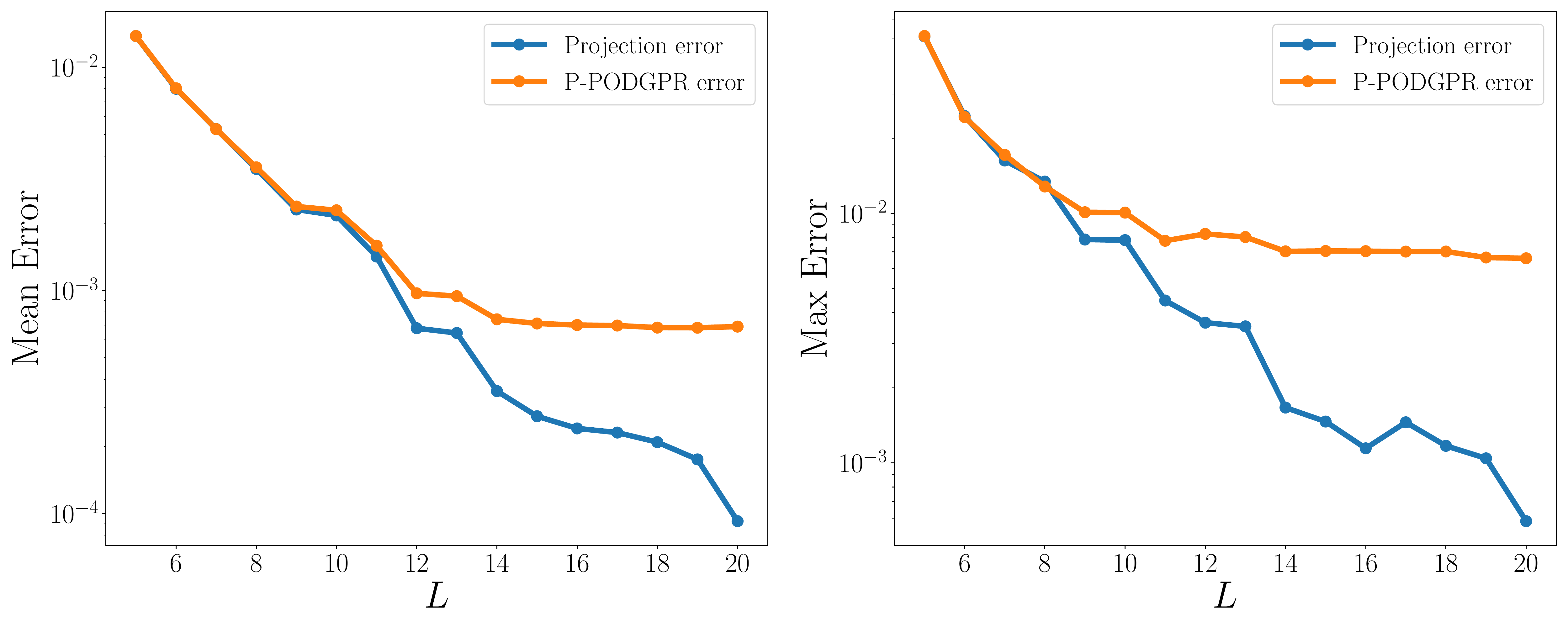}
    \caption{Porous material: Comparison of stress errors for different $L$ with $N_{\rm{pod}}=N_{\rm{reg}}=50$.}
    \label{fig:rve1comparisonL}
\end{figure}

\begin{figure}[p]
    \centering
    \includegraphics[width=0.95\textwidth]{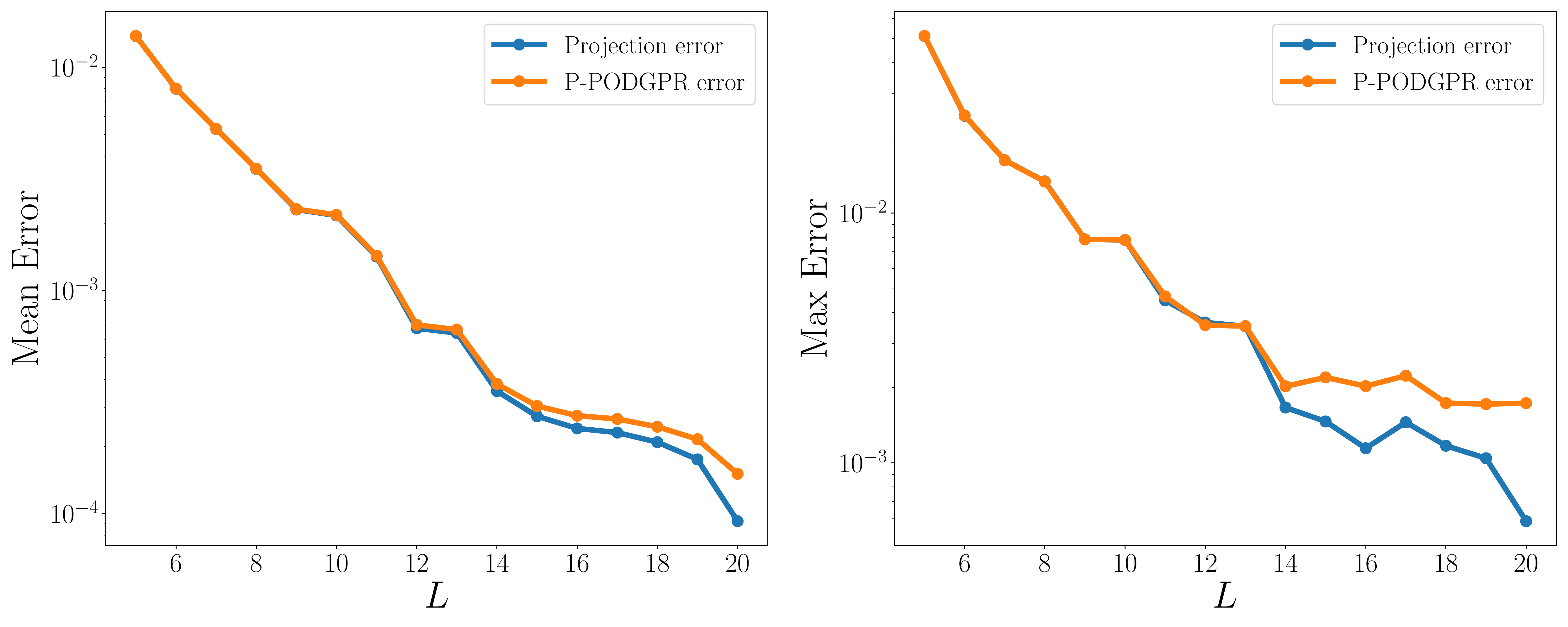}
    \caption{Porous material: Comparison of stress errors for different $L$ with $N_{\rm{pod}}=50$ and $N_{\rm{reg}}=500$.}
    \label{fig:rve1comparisonL500}
\end{figure}

\paragraph{Influence of $L$}
Using $N_{\rm{pod}}=N_{\rm{reg}}=50$, the influence of the number of basis functions $L$ was then investigated. The relative errors of the effective stress are given in Fig.~\ref{fig:rve1comparisonL}. The projection error is also plotted to compare the quality of the regression in Eq.~\eqref{eq:alphamap}. As seen from the figure, both the mean and maximum error of the first 8 basis functions nearly perfectly match. However, for a larger number of basis functions, the discrepancy slowly grows and the error flattens. Generally, the POD coefficients get increasingly more oscillatory with increasing number and hence require more data for an accurate regression.
To show that with increasing $N_{\rm{reg}}$, the regression error decreases and approaches the projection error, the error over $L$ for $N_{\rm{pod}}=50$ with $N_{\rm{reg}}=500$ is also plotted in Fig.~\ref{fig:rve1comparisonL500}.
For this case, the mean error matches the projection error for 13 basis functions, while the maximum error also gets much closer to the projection error. Note that the maximum error of P-PODGPR is sometimes slightly lower than the projection error. This can be explained with errors in the regression that lead to a stress field which after averaging ends up closer to the high fidelity solution than the best projection. This is entirely random and as $N_{\rm{reg}}$ is further increased, the regression error will tend towards zero and the curves will match eventually.

\paragraph{Comparison with Neural Networks}
For comparison, a deep forward neural network with different architectures was trained with the available deformation and stress data collected from the RVE simulations. The stress field data was averaged to give effective stresses. Regarding the network architecture, different neural networks with $N_h=\{1,2\}$ hidden layers with each $N_n=\{20,50\}$ neurons were tested. The input layer and output layer both had 3 and 4 neurons respectively, one for each component of the stretch tensor and effective stress. Apart from the output layer an \textit{ELU} activation function was applied after each layer. Before training, the deformation and stress data were normalized to [0, 1]. Then, all available $N_{\rm{reg}}=500$ training snapshots and all 1000 test snapshots were set as training and validation dataset for the optimization. The network was trained using a mean squared error loss function, the \textit{Adam} optimizer with a learning rate of $1\times10^{-4}$ and a batch size of 32 for 10000 epochs. The validation loss and relative errors $\epsilon_{\bar{\mathbf{P}}}$ obtained for each architecture are given in Tab.~\ref{tab:rve1nn}, where the best architecture is highlighted.
\begin{table}[tb]
    \centering
    \begin{tabular}{l|ccc}
Architecture              & Validation Loss      & $\epsilon_{\bar{\mathbf{P}}}^\text{mean}$ & $\epsilon_{\bar{\mathbf{P}}}^\text{max}$ \\ \hline
$N_h=1, N_n=20$  & $2.33\times 10^{-7}$ & \textbf{0.0016}           & 0.0199                                   \\
$N_h=1, N_n=50$  & $\bm{1.72\times 10^{-7}}$ & 0.0017                      & \textbf{0.0136}                          \\
$N_h=2, N_n=20$  & $2.4\times 10^{-7}$ & 0.0019                                    & 0.0184                                   \\
$N_h=2, N_n=50$  & $1.94\times 10^{-7}$ & 0.0018                        & 0.0166    

\end{tabular}
    \caption{Porous material: Validation and effective stress error for different feedforward neural network architectures. The lowest values in each column have been highlighted.}
    \label{tab:rve1nn}
\end{table}
All architectures produce similar results with the second architecture $N_h=1, N_n=50$ yielding the best results with an average and maximum error of 0.17\% and 1.36\%. On the other hand, P-PODGPR already yields a lower mean and maximum error of around 0.065\% and 0.65\% when using only 50 training data, showing that P-PODGPR can utilize the information of each snapshot more efficiently than the neural network. When 500 snapshots are used for the regression in P-PODGPR, a mean and max error of less than 0.02\% and 0.2\% is achieved, hence outperforming the neural network greatly.

\subsubsection{Fiber reinforced material}
\label{subsubsec:composite}
\begin{figure}[tb]
    \centering
    \includegraphics[width=0.3\textwidth]{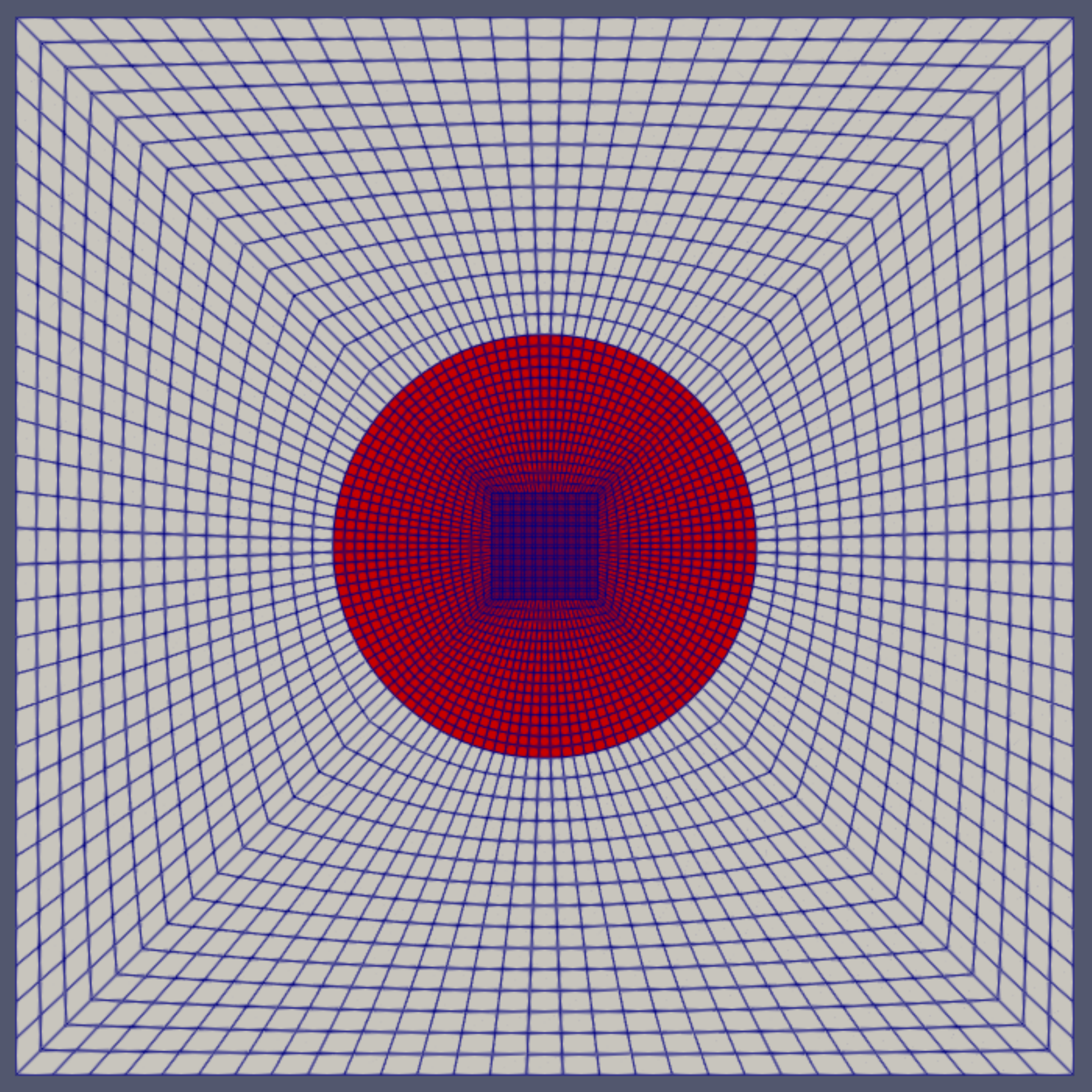}
    \caption{Fiber reinforced material: The considered microstructure consisted of a Neo-Hookean matrix material and a Neo-Hookean fiber material. The fiber volume fraction was 12.56\%. The mesh consisted of 4321 eight-node elements and 13080 nodes.}
    \label{fig:rve2mesh}
\end{figure}

In the second example, the considered microstructural RVE consists of two different phases: a soft matrix and a stiff fiber material. Both materials are Neo-Hookean, but the matrix has parameters $C_1=1$, $D_1=1$, while the fiber has variable parameters $C_1=D_1\in[50, 150]$, corresponding to a Young's modulus that is 50-150 times higher than the matrix, with the Poisson ratio remaining the same ($\nu=0.25$), cf. Eq.~\eqref{eq:lameconstants}. The geometry used is depicted in Fig.~\ref{fig:rve2mesh}, where the matrix (grey) completely surrounds the fiber (red). Eight node quadrilateral elements with four quadrature points are employed. The volume fraction of the fiber is 12.56\%. Linear displacement boundaries with $\bar{\bm{U}}-\mathbf{I}\in [-0.3,0.3]$ and fiber parameters of $C_1=D_1\in[50, 150]$ are considered. Therefore, the considered problem has 4 varying parameters.

\paragraph{Data generation}
A set of 1000 training snapshots and 1000 test snapshots were generated. The first set contained the $2^4=16$ corner points of the parameter domain and the remaining points were sampled from a Sobol sequence, while the second set was generated from a random uniform distribution.

\begin{figure}[tb]
    \centering
    \includegraphics[width=0.8\textwidth]{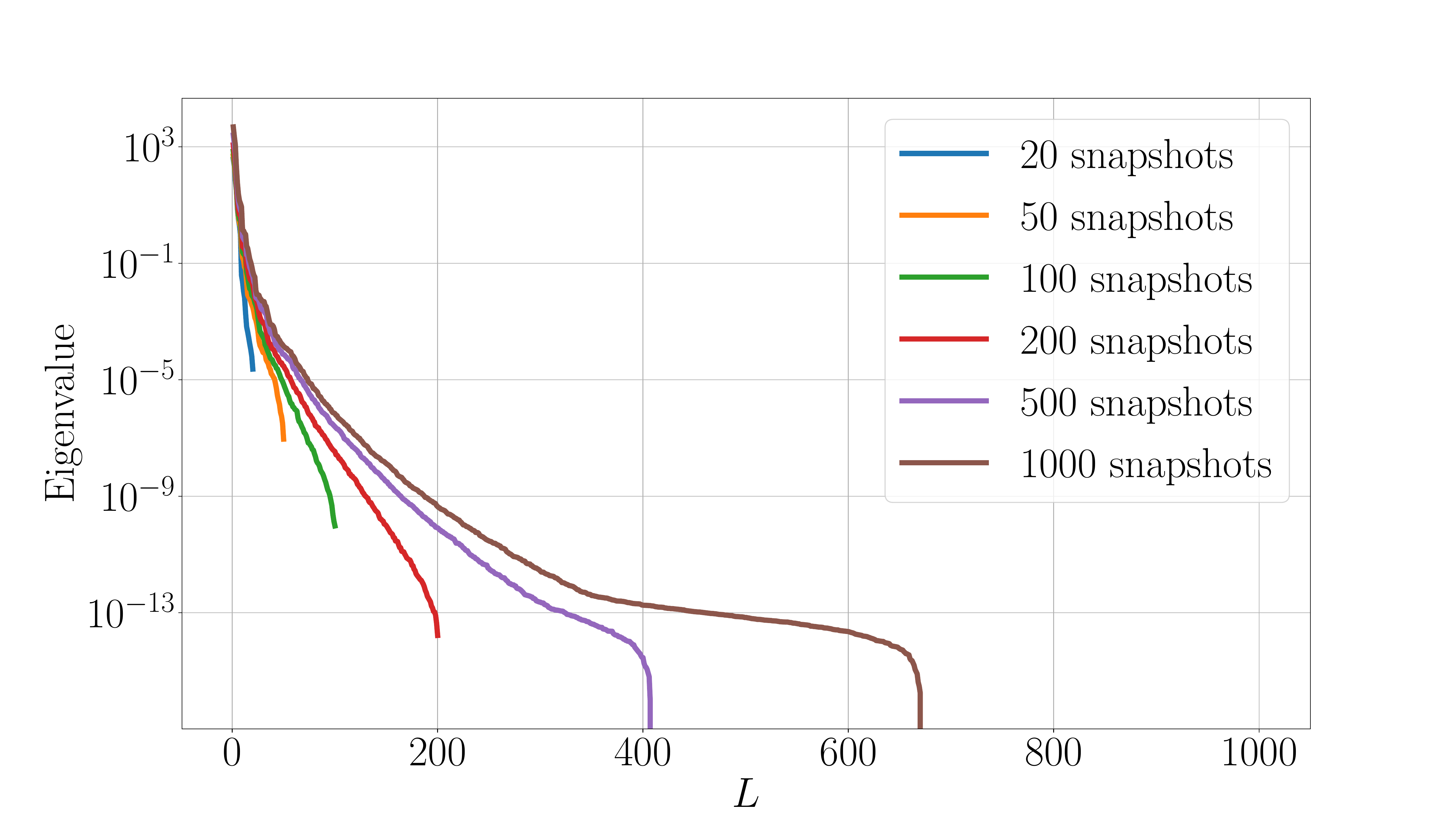}
    \caption{Fiber reinforced material: Eigenvalues of the correlation matrix for different numbers of snapshots used for POD.}
    \label{fig:rve2eig}
\end{figure}

\begin{figure}[tb]
     \centering
     \begin{subfigure}[b]{0.3\textwidth}
         \centering
         \includegraphics[width=\textwidth]{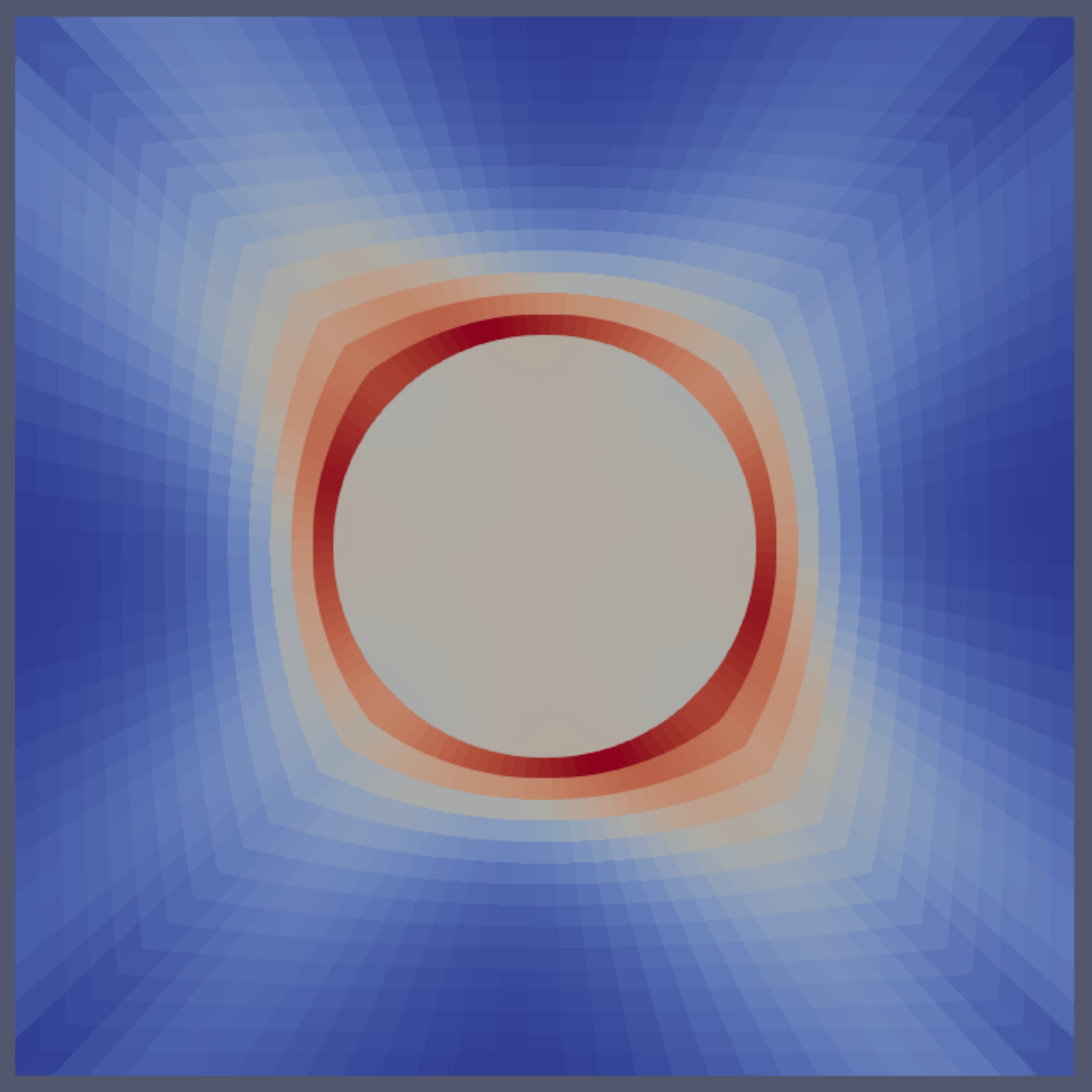}
         \caption{1st basis function}
         \label{fig:rve2basisfunc1}
     \end{subfigure}
     \hfill
     \begin{subfigure}[b]{0.3\textwidth}
         \centering
         \includegraphics[width=\textwidth]{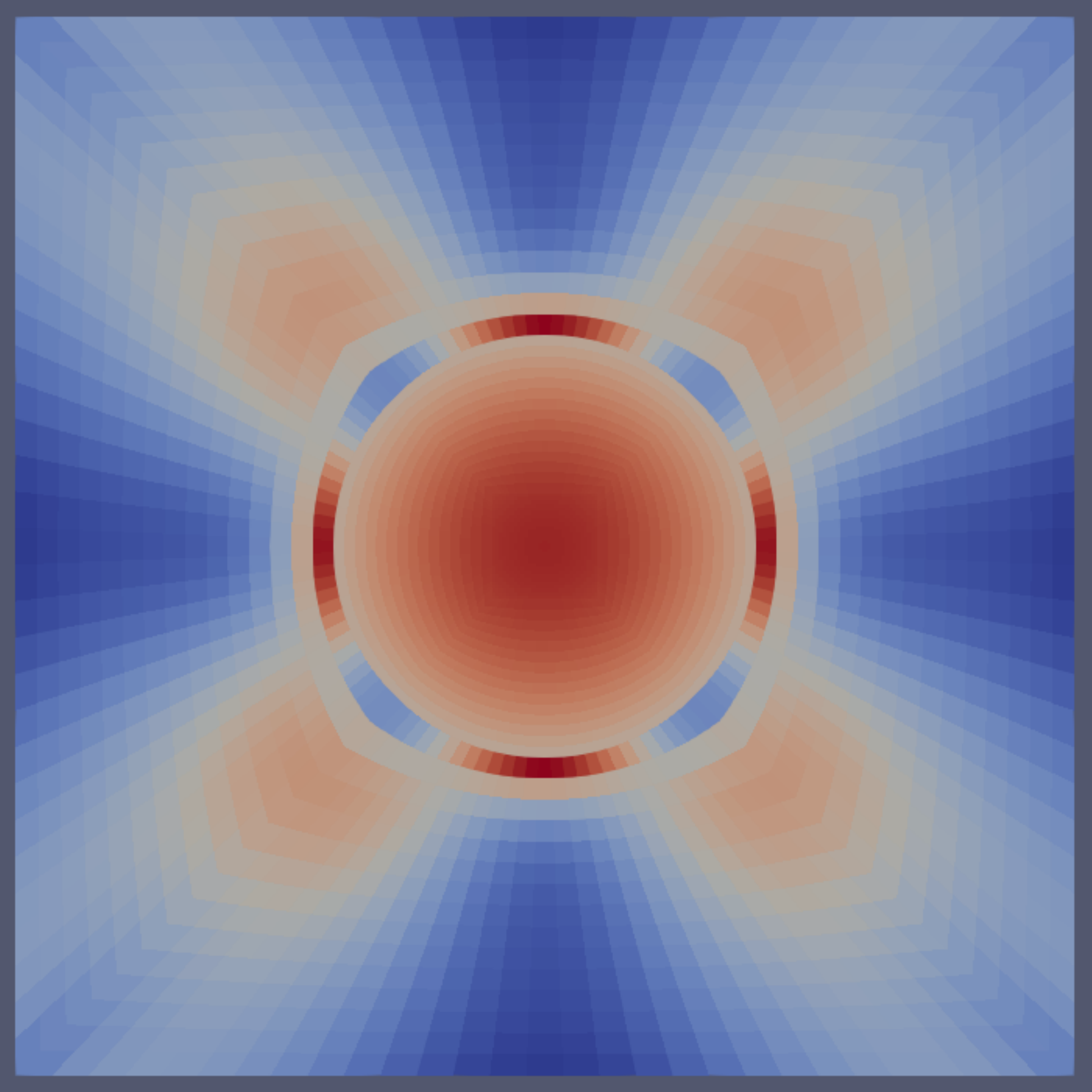}
         \caption{2nd basis function}
         \label{fig:rve2basisfunc2}
     \end{subfigure}
     \hfill
     \begin{subfigure}[b]{0.3\textwidth}
         \centering
         \includegraphics[width=\textwidth]{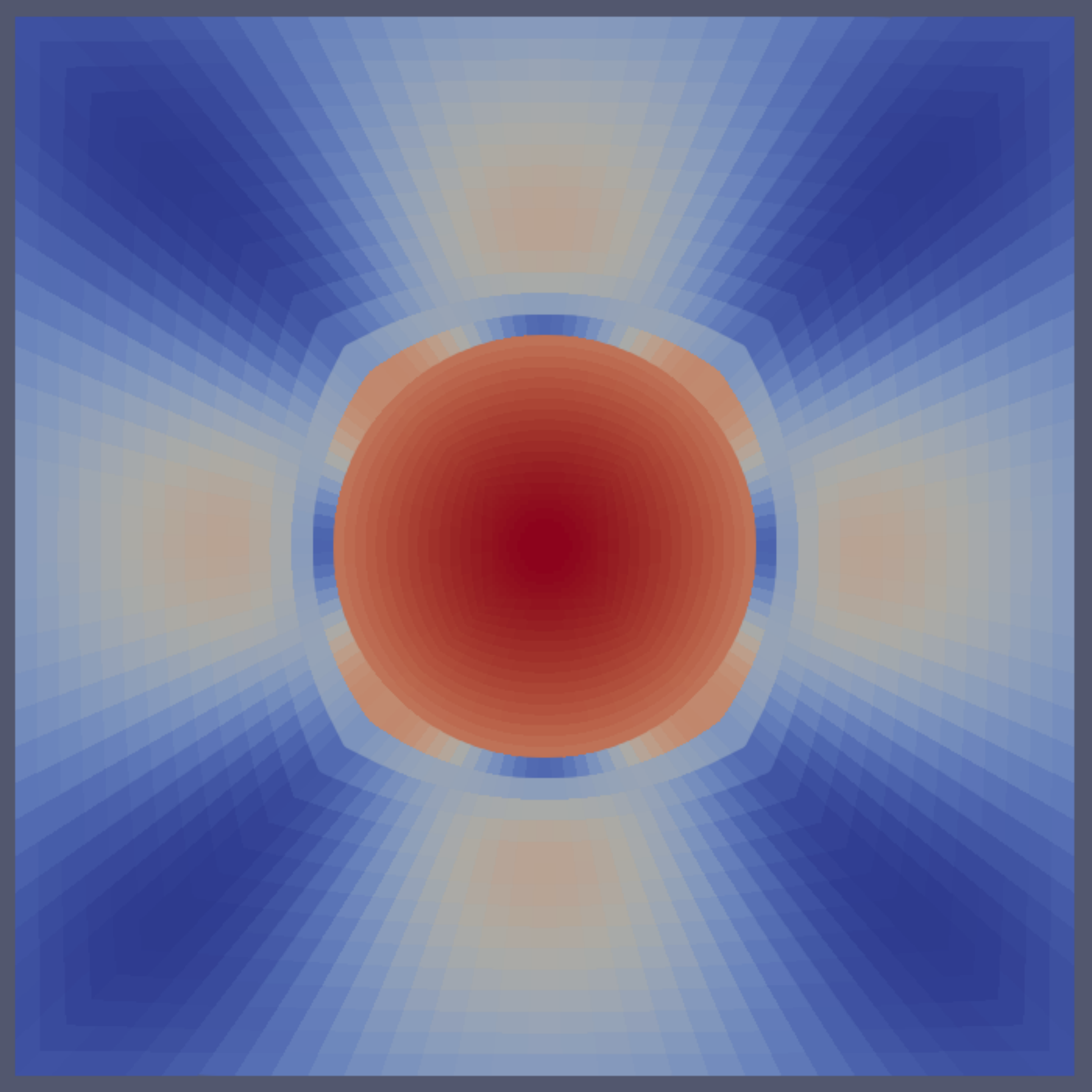}
         \caption{3rd basis function}
         \label{fig:rve2basisfunc3}
     \end{subfigure}
    \caption{Fiber reinforced material: Von Mises stress of the first three POD basis functions of the microscopic stress field.}
    \label{fig:rve2basisfunc}
\end{figure}

\paragraph{Eigenvalues}
The eigenvalues of the correlation matrix for different numbers of training snapshots are plotted in Fig.~\ref{fig:rve2eig}. Similar to the last example, for all cases, an exponential decay can be observed, showing the reducibility of the problem. 
The von Mises stress of the first three POD basis functions are plotted in Fig.~\ref{fig:rve2basisfunc}. Subsequently, $L=20$ basis functions are considered which correspond to an energy $\mathcal{E}_{\rm{pod}}$ of $99.9901\%$. 

\begin{figure}[p]
    \centering
    \includegraphics[width=0.95\textwidth]{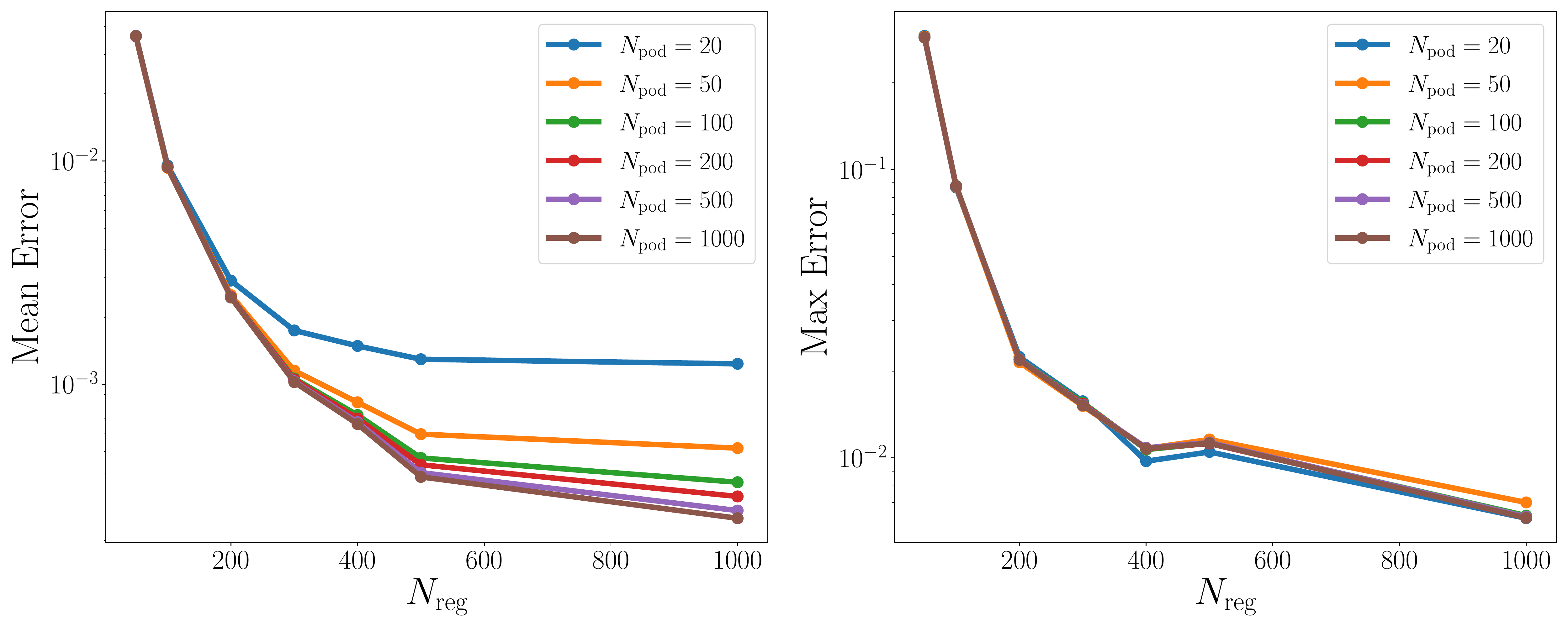}
    \caption{Fiber reinforced material: Comparison of stress errors for different combinations of $N_{\rm{pod}}$ and $N_{\rm{reg}}$ for $L=20$.}
    \label{fig:rve2comparison}
\end{figure}

\paragraph{Influence of $N_{\rm{pod}}$ and $N_{\rm{reg}}$}
For $L=20$ basis functions, combinations of $N_{\rm{pod}}\in\{20,50,100,200,500\}$ and $N_{\rm{reg}}\in\{100,200,300,400,500,1000\}$ were tested. The mean and maximum error plots can be seen in Fig.~\ref{fig:rve2comparison}.
It can be observed that for all cases with $N_{\rm{reg}}>50$, the mean error is below 1\%. The mean error improves as more snapshots were considered for the POD basis, suggesting that there was still new information in the snapshots, which fine-tune the optimal basis. The mean error reaches around 0.04\% for $N_{\rm{reg}}=500$ snapshots with a maximum error of around 1\%. Increasing $N_{\rm{reg}}$ from 500 to 1000 does not improve the mean error significantly, meaning that the projection error has already been reached with 500 snapshots and $L=20$ basis functions. Weighing the approximation error over the number of full computations needed, $N_{\rm{pod}}=N_{\rm{reg}}=200$ is considered in the subsequent analysis.

\begin{figure}[p]
     \centering
     \includegraphics[width=0.95\textwidth]{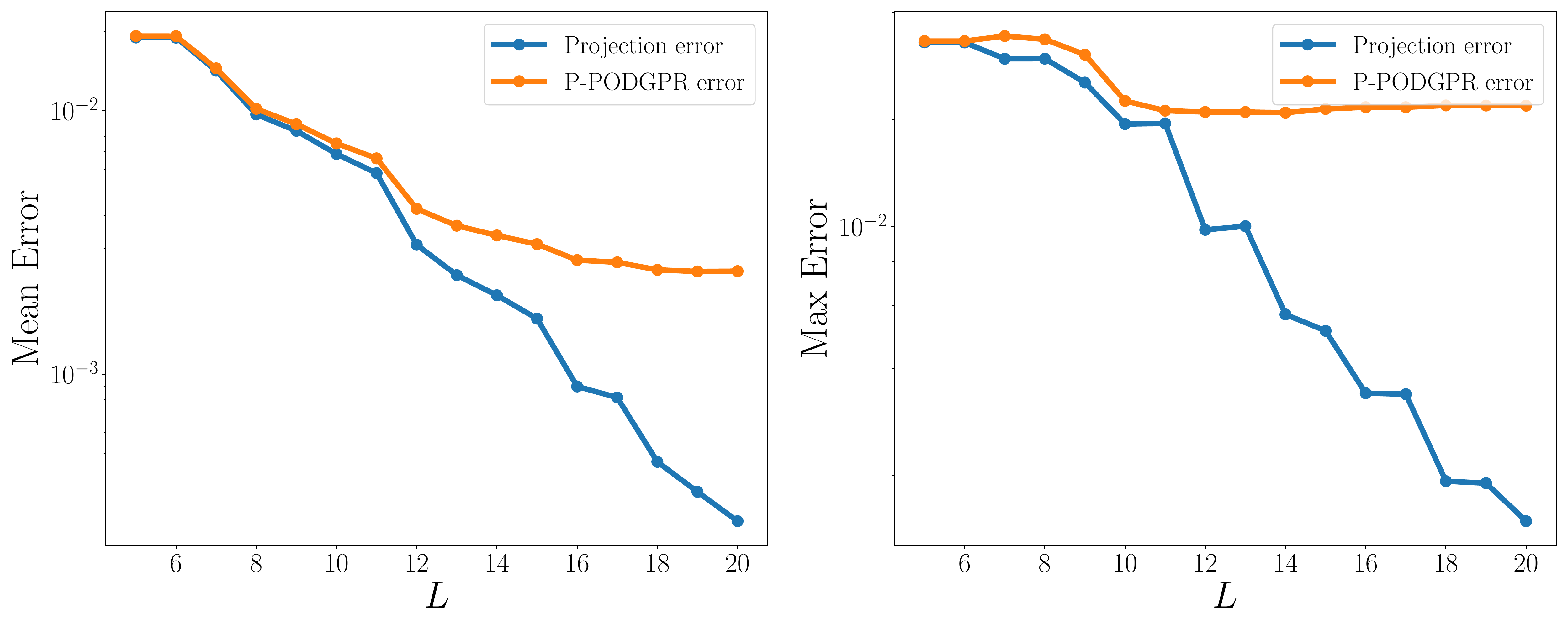}
        \caption{Fiber reinforced material: Comparison of stress errors for different $L$ with $N_{\rm{pod}}=200$ and $N_{\rm{reg}}=200$.}
        \label{fig:rve2comparisonL}
\end{figure}

\paragraph{Influence of $L$}
Using $N_{\rm{pod}}=N_{\rm{reg}}=200$, the influence of $L$ is investigated. The mean and maximum error of the approximation error over the number of basis functions $L$ used are shown in Fig.~\ref{fig:rve2comparisonL}. The projection error is also plotted to reveal the quality of the regression of Eq.~\eqref{eq:alphamap}.
\begin{figure}[p]
     \centering
     \includegraphics[width=0.95\textwidth]{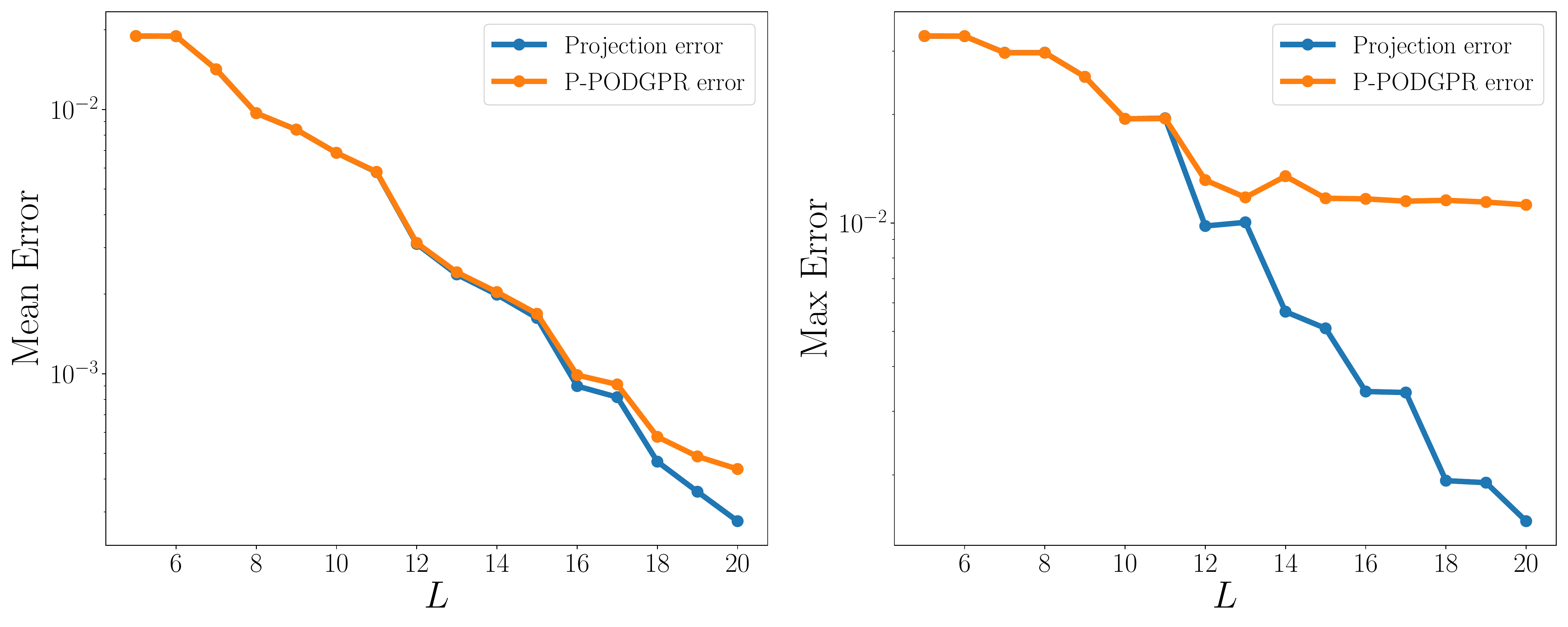}
        \caption{Fiber reinforced material: Comparison of stress errors for different $L$ with $N_{\rm{pod}}=200$ and $N_{\rm{reg}}=500$.}
        \label{fig:rve2comparisonL500}
\end{figure}
As seen from the figure, the mean and maximum error for the first 11 basis functions match the projection error quite well, and afterwards flatten out, similar to the last example. When the number of snapshots $N_{\rm{reg}}$ is increased to 500, the mean error matches the projection error for 16 basis functions, as seen in Fig.~\ref{fig:rve2comparisonL500}, while the maximum error also slightly improves.

\paragraph{Comparison with Neural Network}
Similarly to the last example, deep forward neural networks with different architectures were trained for comparison. The available 500 training data and 1000 testing data were used for training and validation. The same network architecture configurations as before were tested, with one additional combination $N_h=2, N_n=100$. The results are given in Tab.~\ref{tab:rve2nn}.
\begin{table}[tb]
    \centering
    \begin{tabular}{l|ccc}
Architecture              & Validation Loss               & $\epsilon_{\bar{\mathbf{P}}}^\text{mean}$ & $\epsilon_{\bar{\mathbf{P}}}^\text{max}$ \\ \hline
$N_h=1, N_n=20$  & $1.7\times 10^{-6}$          & 0.0077                                    & 0.0368                                   \\
$N_h=1, N_n=50$  & $8.55\times 10^{-7}$          & 0.0056                                    & 0.0289
                                   \\
$N_h=2, N_n=20$  & $5.36\times 10^{-7}$          & 0.0047                           & \textbf{0.0176}                                   \\
$N_h=2, N_n=50$  & $\bm{2.97\times 10^{-7}}$     & \textbf{0.0039}                           & 0.0206  \\
$ N_h=2, N_n=100$ & $7.91\times 10^{-7}$  & 0.0052  & 0.0315
\end{tabular}
    \caption{Fiber reinforced material: Validation loss and effective stress error for different feedforward neural network architectures. The lowest values in each column have been highlighted in bold face.}
    \label{tab:rve2nn}
\end{table}
The fourth architecture $N_h=2, N_n=50$ performs the best with an average error of 0.39\% and a maximum error of 2.06\%. Same as for the previous example, P-PODGPR outperforms the neural network. With only 200 training snapshots, P-PODGPR achieves roughly the same accuracy as the NN using 500 snapshots. With 500 employed snapshots, P-PODGPR reaches a mean and maximum error of 0.04\% and 1\%, hence outperforming the neural network, while also being able to recover the microscopic stress field.

\begin{figure}[tb]
     \centering
     \includegraphics[width=0.95\textwidth]{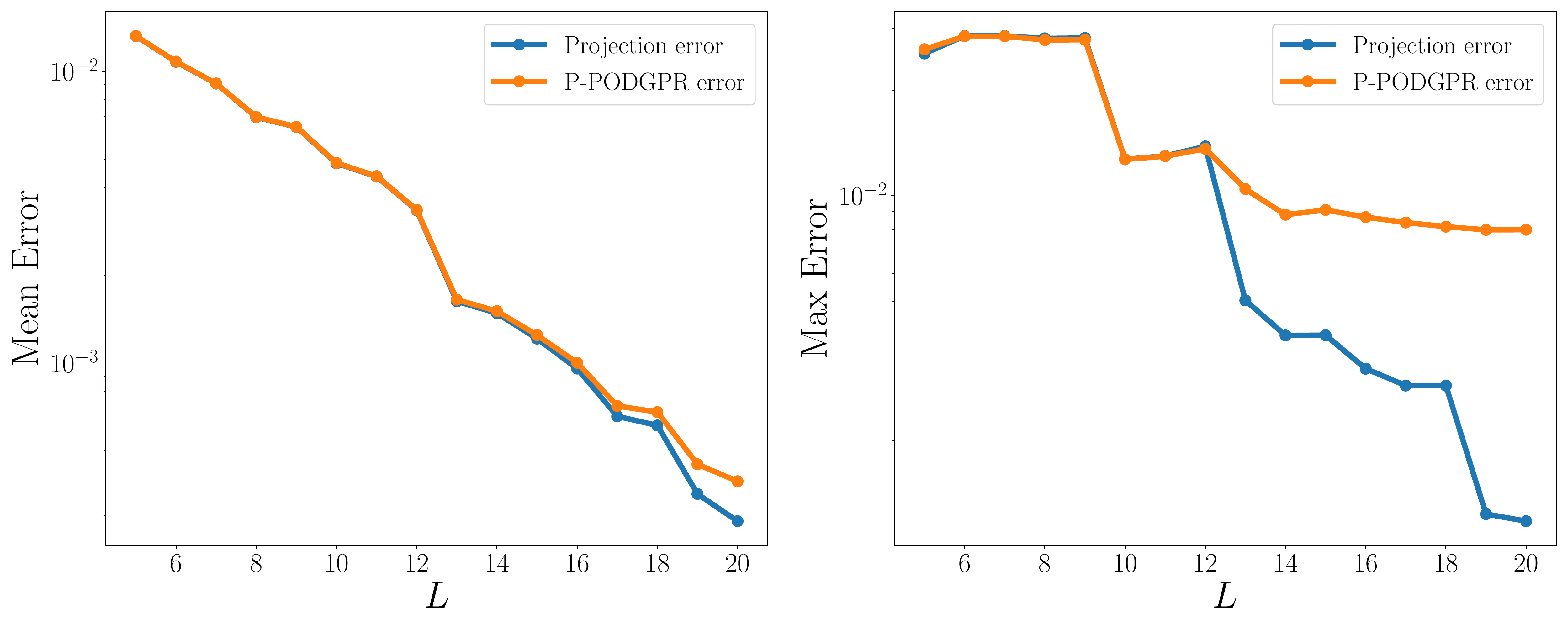}
        \caption{Fiber reinforced material with periodic boundary conditions: Comparison of stress errors for different $L$ with $N_{\rm{pod}}=200$ and $N_{\rm{reg}}=500$.}
        \label{fig:rve2comparisonL500_pbc}
\end{figure}

\paragraph{Remark on Periodic Boundary Conditions}
For the same microstructure (Fig.~\ref{fig:rve2mesh}), an analogous analysis using periodic boundary conditions was considered. In contrast to linear boundary conditions, the fluctuation displacement field $\bm{w}$ is assumed to be periodic on the RVE boundary. The obtained results regarding the approximation quality are shown in Fig.~\ref{fig:rve2comparisonL500_pbc}, where $N_{\rm{pod}}=200$ and $N_{\rm{reg}}=500$ was chosen. In comparison to Fig.~\ref{fig:rve2comparisonL500}, the results are comparable and hence it can be concluded that P-PODGPR works independently of the chosen boundary conditions.

\subsection{Two-scale simulation}
\begin{figure}[tb]
    \centering
    \includegraphics[width=0.8\textwidth]{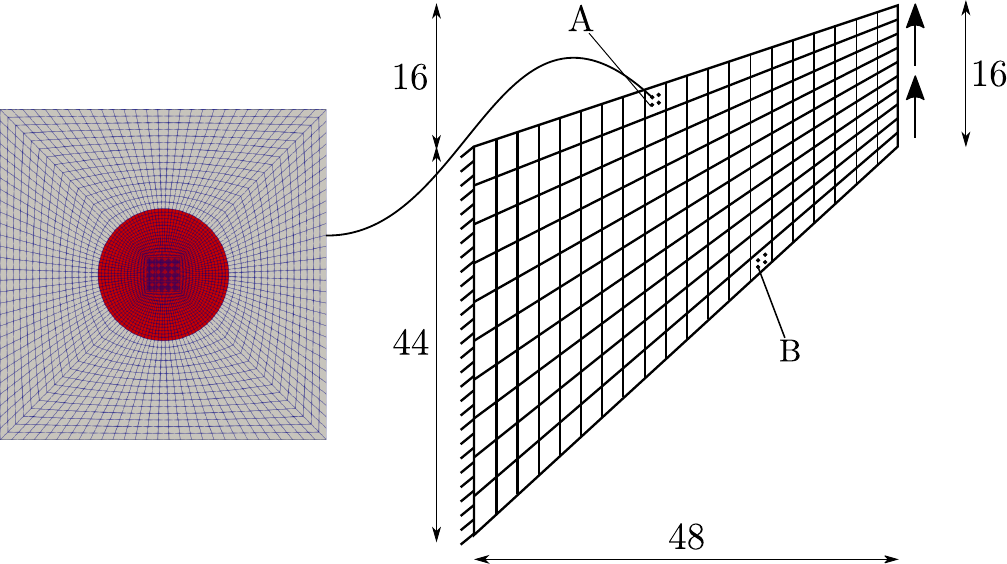}
    \caption{Cook's membrane: The left side was fixed while a traction in the vertical direction of $0.1$ was applied along the right side. The mesh consisted of 200 bilinear elements and 231 nodes. The microstructural stress field at the marked quadrature points A and B will be shown later.}
    \label{fig:macromesh}
\end{figure}

To show the performance of P-PODGPR in a two-scale simulation, the learned constitutive model for the fiber reinforced RVE with $N_{\rm{pod}}=N_{\rm{reg}}=200$ is embedded inside a FE solver. For the macroscopic problem the Cook's membrane, consisting of the fiber reinforced RVE, is chosen. The geometry and mesh of the membrane are given in Fig.~\ref{fig:macromesh}. The macroscopic mesh consists of 200 bilinear quadrilateral elements with 4 quadrature points. The quadrature points A and B, as marked on the figure, denote points in which the microscopic stress field is compared against the reference solution. The reference solution is obtained by running a full two-scale FE$^2$ simulation. The left side of the sample is fixed and the right side is loaded in five time steps with a vertical traction of $0.1$. The material parameters of the fiber are taken to be fixed with $C_1=D_1=100$ in this example.

\begin{figure}[p]
     \centering
     \begin{subfigure}[b]{0.3\textwidth}
         \centering
         \includegraphics[width=\textwidth]{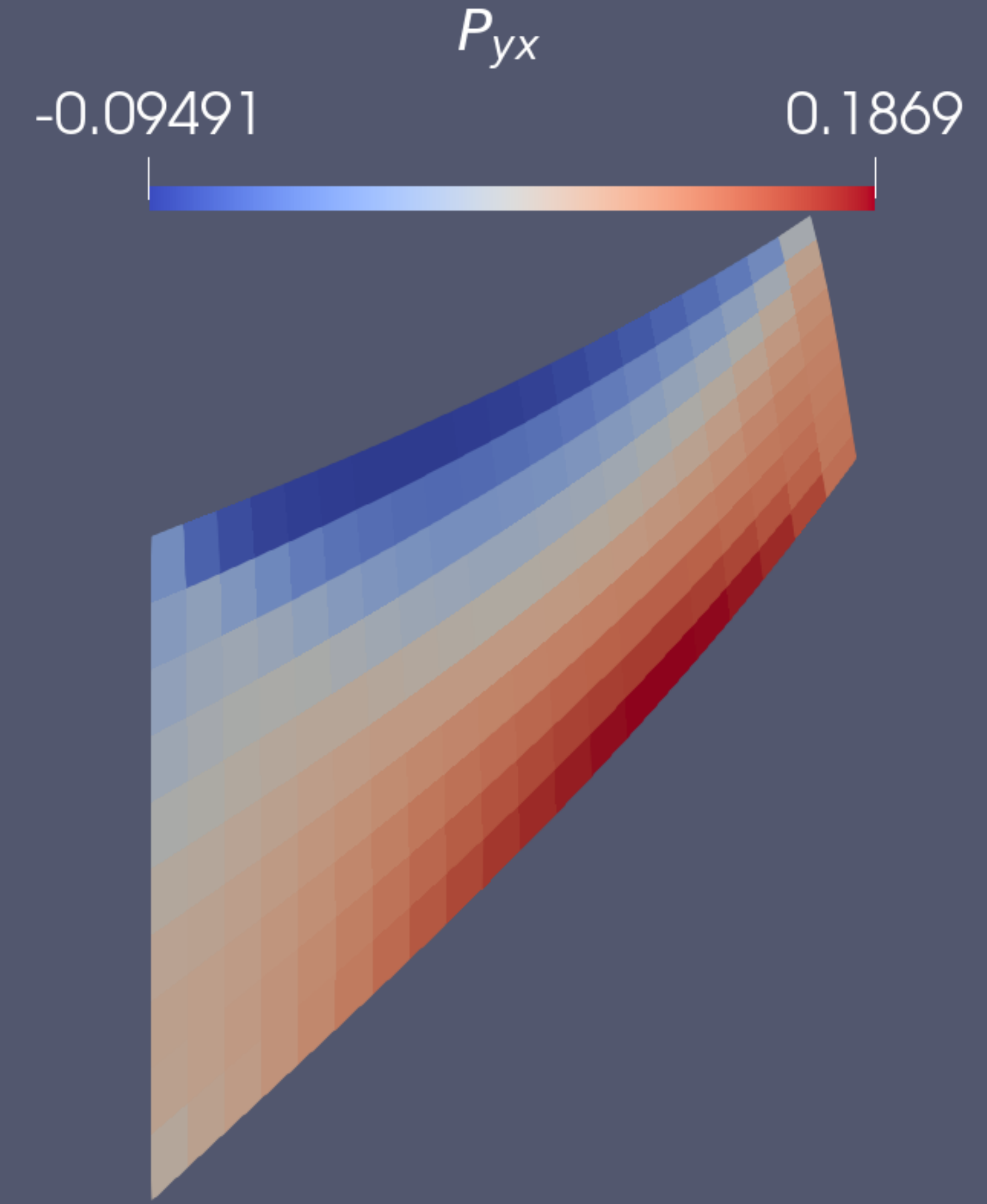}
         \caption{FE$^2$}
         \label{fig:macro_fe2}
     \end{subfigure}
     \hfill
     \begin{subfigure}[b]{0.3\textwidth}
         \centering
         \includegraphics[width=\textwidth]{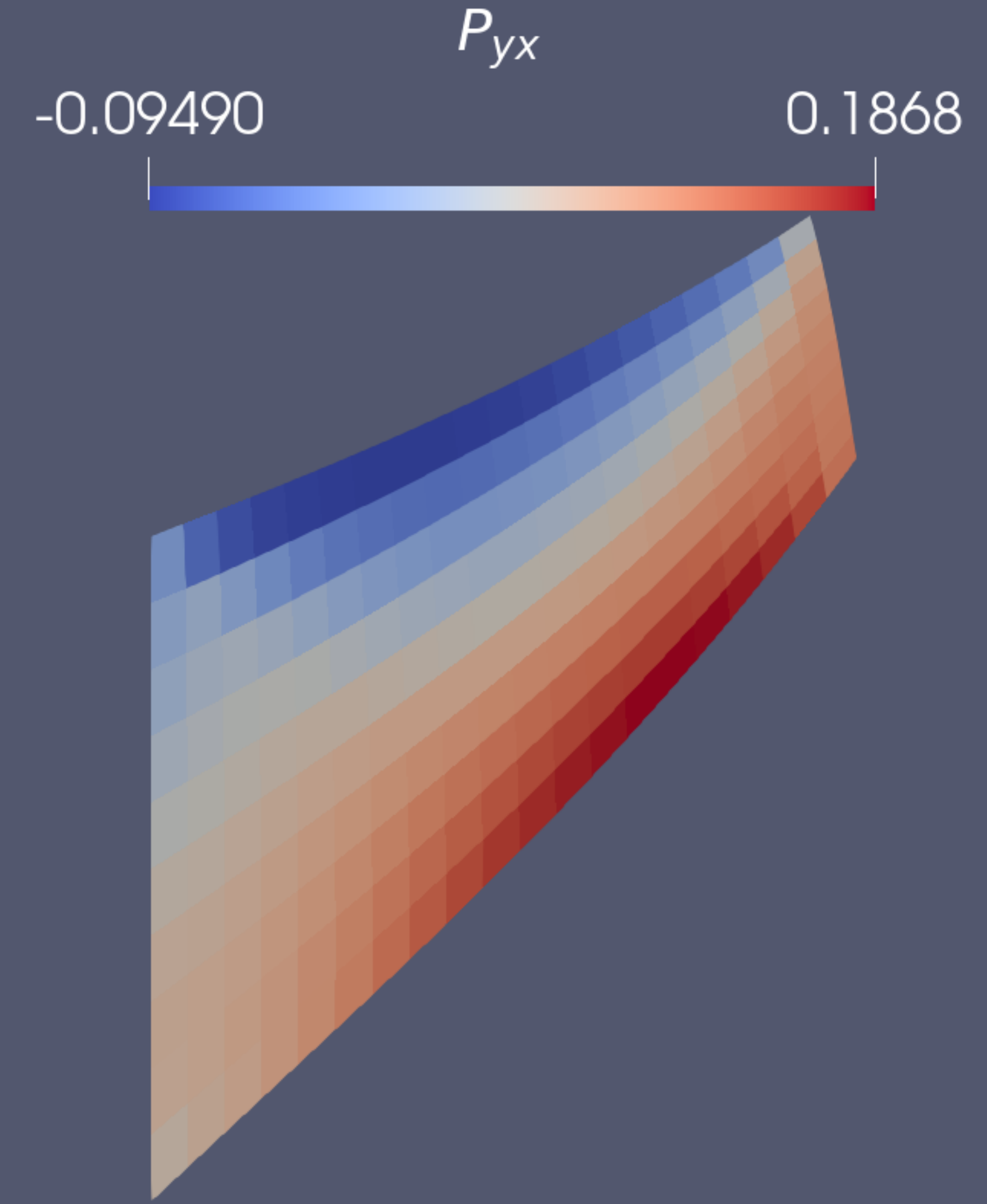}
         \caption{FE with P-PODGPR}
         \label{fig:macro_rom}
     \end{subfigure}
     \hfill
     \begin{subfigure}[b]{0.3\textwidth}
         \centering
         \includegraphics[width=\textwidth]{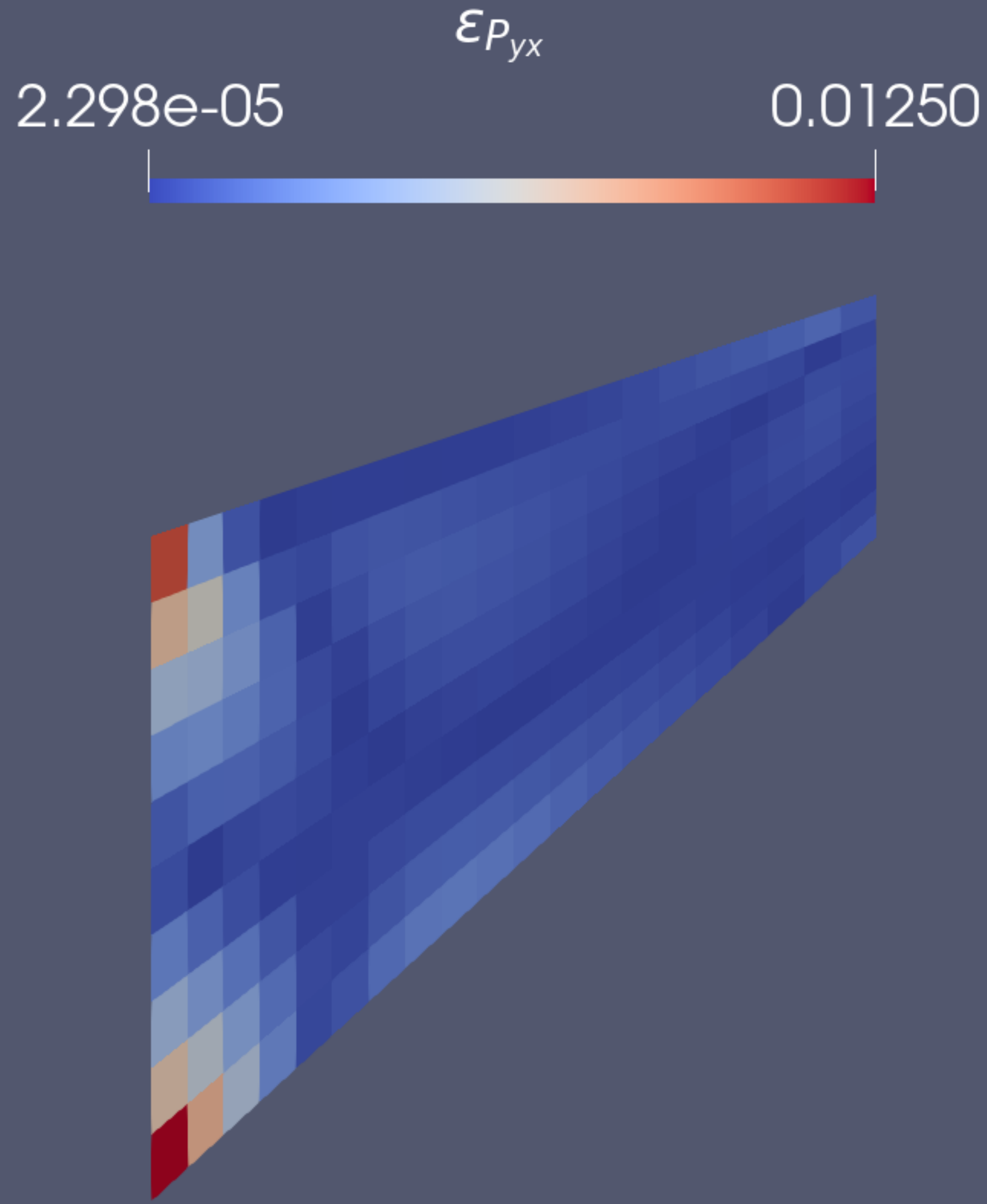}
         \caption{Relative error}
         \label{fig:errorMacro}
     \end{subfigure}
        \caption{Cook's membrane: Macroscopic stress field.}
        \label{fig:fe2comparison}
\end{figure}

\begin{figure}[p]
     \centering
     \begin{subfigure}[b]{0.3\textwidth}
         \centering
         \includegraphics[width=\textwidth]{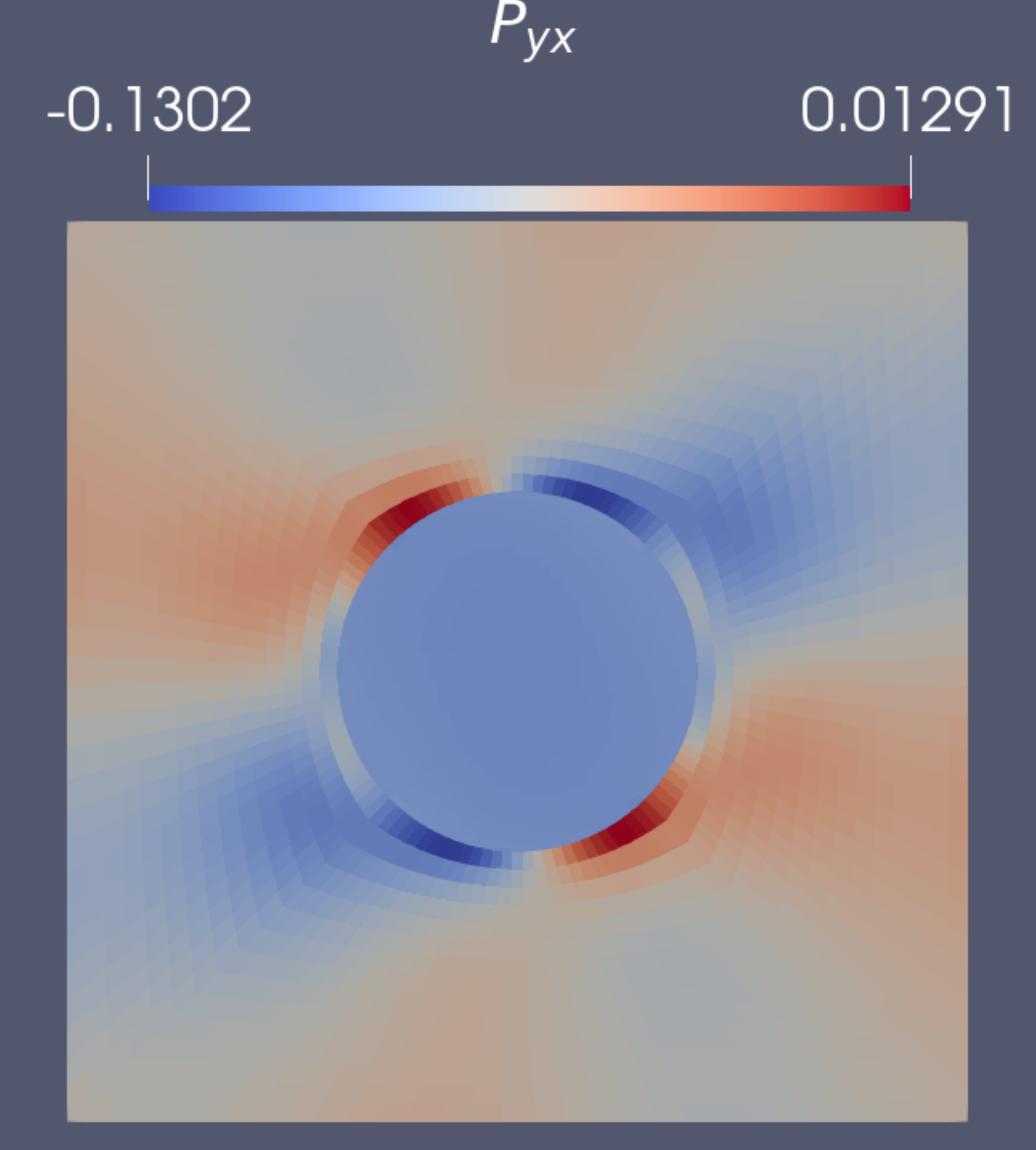}
         \caption{FE$^2$}
         \label{fig:microA_FE2}
     \end{subfigure}
     \hfill
     \begin{subfigure}[b]{0.3\textwidth}
         \centering
         \includegraphics[width=\textwidth]{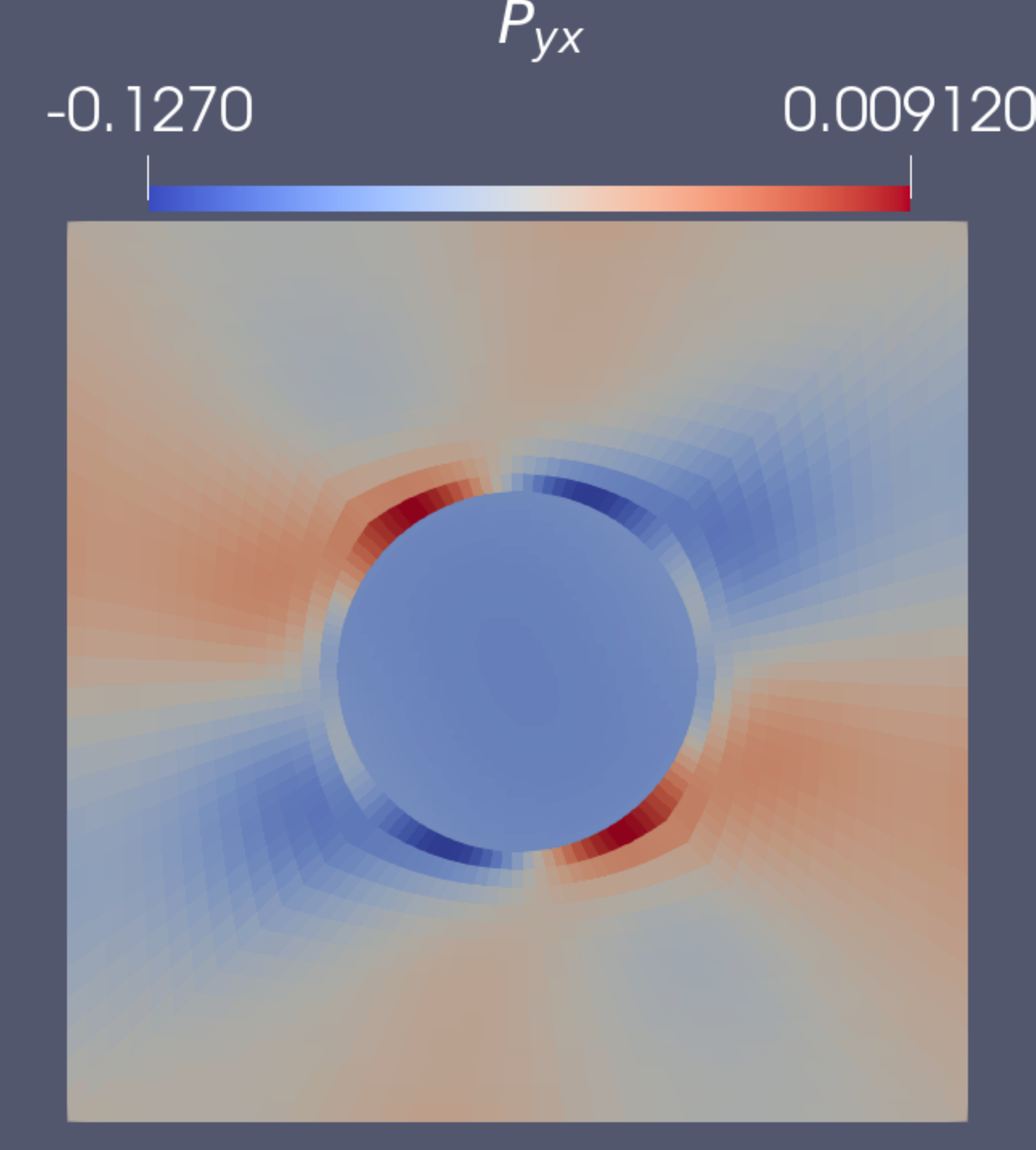}
         \caption{FE with P-PODGPR}
         \label{fig:microA_ROM}
     \end{subfigure}
     \hfill
     \begin{subfigure}[b]{0.3\textwidth}
         \centering
         \includegraphics[width=\textwidth]{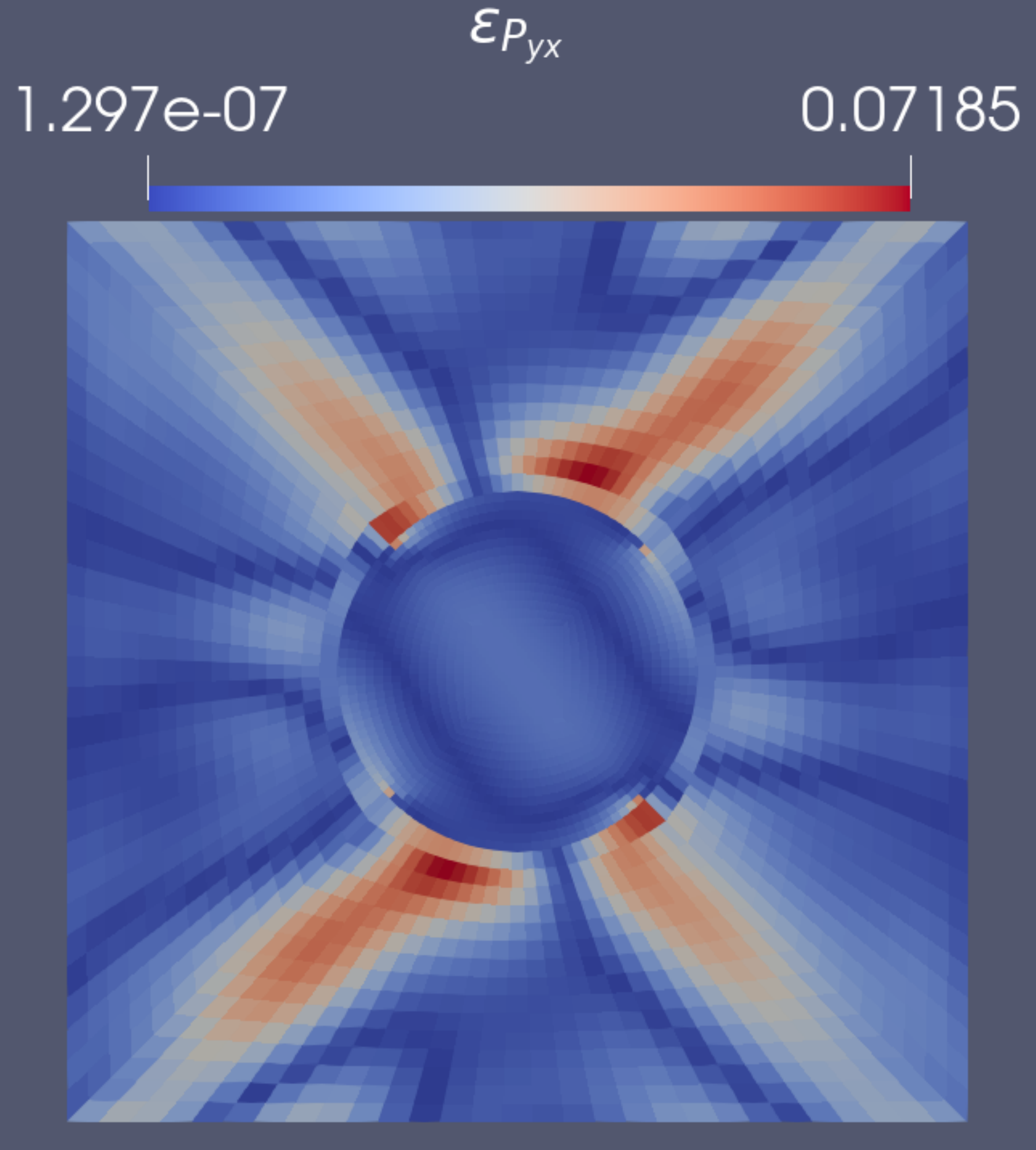}
         \caption{Relative error}
         \label{fig:error_A}
     \end{subfigure}
        \caption{Cook's membrane: Microscopic stress field at Point A.}
        \label{fig:microA}
\end{figure}

\begin{figure}[p]
     \centering
     \begin{subfigure}[b]{0.3\textwidth}
         \centering
         \includegraphics[width=\textwidth]{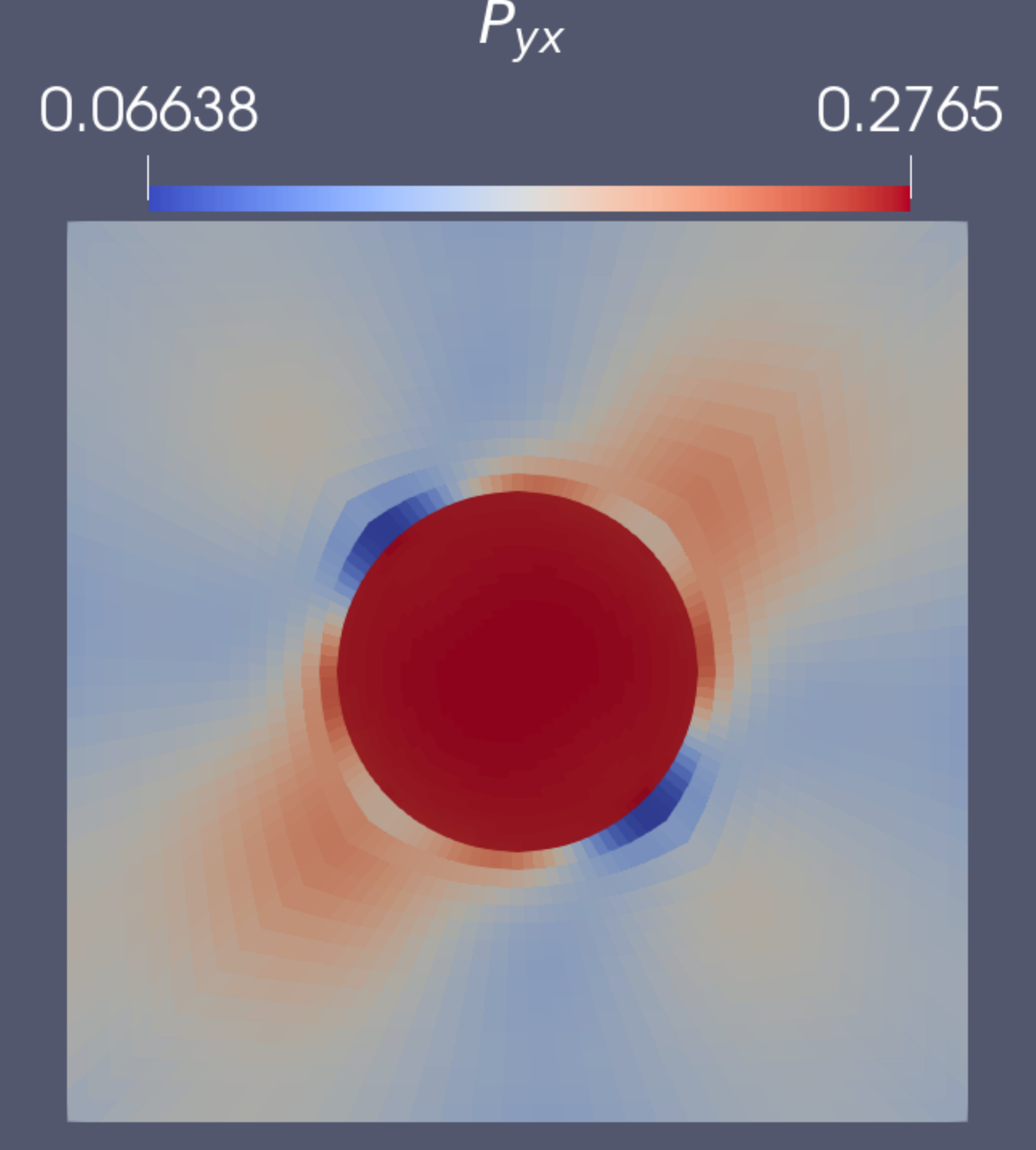}
         \caption{FE$^2$}
         \label{fig:microB_FE2}
     \end{subfigure}
     \hfill
     \begin{subfigure}[b]{0.3\textwidth}
         \centering
         \includegraphics[width=\textwidth]{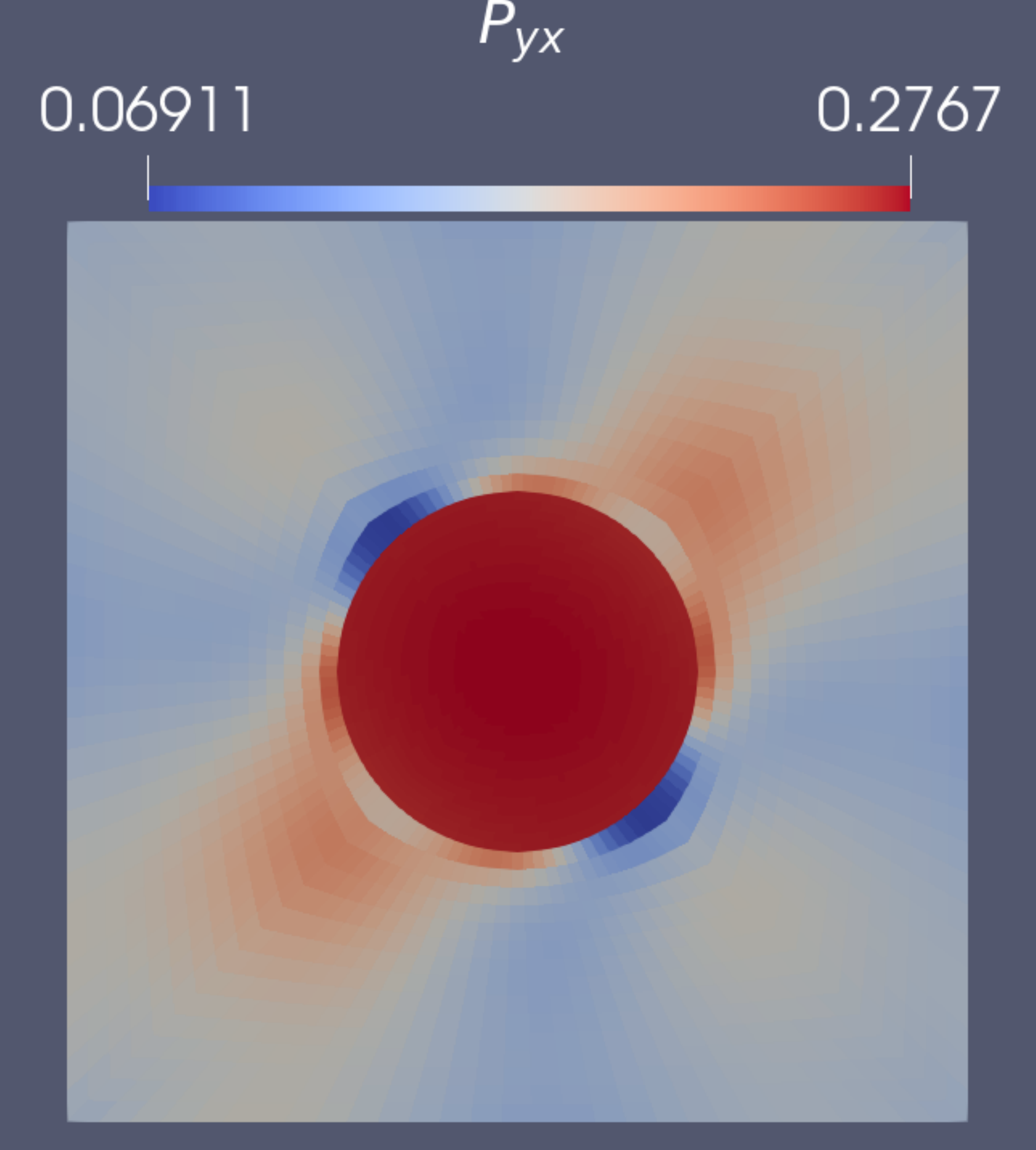}
         \caption{FE with P-PODGPR}
         \label{fig:microB_ROM}
     \end{subfigure}
     \hfill
     \begin{subfigure}[b]{0.3\textwidth}
         \centering
         \includegraphics[width=\textwidth]{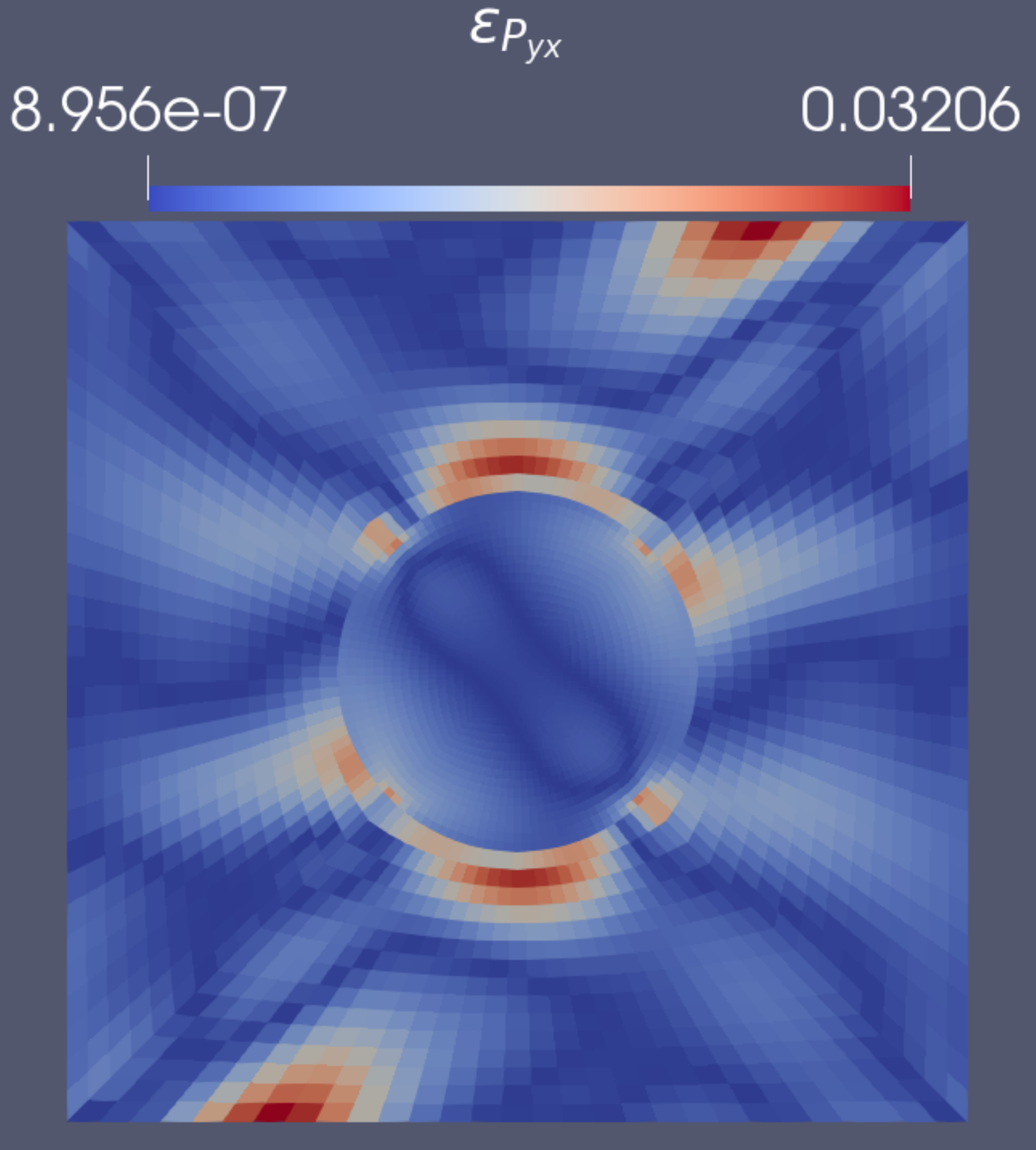}
         \caption{Relative error}
         \label{fig:error_B}
     \end{subfigure}
        \caption{Cook's membrane: Microscopic stress field at Point B.}
        \label{fig:microB}
\end{figure}

The $yx$-component of the macroscopic stress $\bar{P}_{yx}$ obtained by the full FE$^2$ and FE with P-PODGPR simulation are given in Fig.~\ref{fig:fe2comparison}, while the microscopic stress $P_{yx}$ at point A and B are shown in Fig.~\ref{fig:microA} and Fig.~\ref{fig:microB} respectively. The relative error defined as $\epsilon_{P_{yx}}\coloneqq |P_{yx}^{\text{FE2}}-P_{yx}^{\text{ROM}}|/\left<|P_{yx}^{\text{FE2}}|\right>$, with $\left<|P_{yx}^{\text{FE2}}|\right>$ the averaged absolute stress, is also shown (Fig.~\ref{fig:errorMacro}, \ref{fig:error_A}, \ref{fig:error_B}). As can be seen, the shape of the stress field of both solutions is indistinguishable. Even though the relative errors for the microscopic problem reaches a maximum of 7\% near the interface of matrix and fiber, after homogenization the highest error reduces to merely 1\%. This discrepancy is due to the fact that the method tries to reduce the $L_2$-norm of the error in the stress field and therefore allows locally high errors.

Using 48 cores\footnote{Intel Xeon Platinum 8260}, the FE$^2$ simulation takes around 100 minutes while the simulation with P-PODGPR is completed within 1 minute on a single core\footnote{Intel Core i7-8750H}, resulting in a speedup of about three orders of magnitude. For P-PODGPR, 200 RVE simulations have to be pre-computed, which takes less than one hour on a single core. Performing the POD and GPR to construct P-PODGPR takes around one minute.

\section{Conclusions}
\label{sec:conclusion}
In this work, the proper orthogonal decomposition (POD) has been applied on the microscopic stress field to find a reduced basis. A direct mapping from the loading and material parameters to the POD basis coefficients was built with Gaussian process regression (GPR). For the two considered microstructures, involving a porous and a fiber-reinforced material, the proposed method captured the stress field accurately with a mean error of 0.1\%, even in the case of varying material parameters. Finally, the learned constitutive model was employed inside a macroscopic simulation and compared to a full scale FE$^2$ simulation, reaching high accuracy on both macro- and microscale, while gaining a speedup of the order of $10^3$.

Although the examples presented here are two dimensional, the theory has been derived for three dimensions and can readily be applied to 3D problems. Due to the non-intrusive nature of the method, it can be easily implemented into any existing finite element solver. It is noted that the framework is not limited to POD and GPR, but in fact any method for discovering the reduced basis and any regression method can be employed.

This novel method has the potential to open up new ways of material design, as the construction of the surrogate model only requires a relatively small dataset and the surrogate can cover a large range of microstructural parameters. Since the derivatives with respect to the parameters are at hand, it is furthermore possible to define and solve macroscopic optimization problems. Moreover, the microscopic stress field can be fully recovered and visualized in order to make informed modifications of the microstructure to relieve local stress concentrations and remove flaws. Currently, works on more complicated microstructures are being conducted.
\section*{Data availability}
The data that support the findings of this study are available from the corresponding author upon request.
\section*{Acknowledgements}
This result is part of a project that has received funding from the European
Research Council (ERC) under the European Union’s Horizon 2020 Research and Innovation
Programme (Grant Agreement No. 818473).

\bibliography{references}
\end{document}